# Reconfigurable 2D Lateral Heterostructure Platform for Unified Detection, Emission, and Logic

Purbasha Ray[1], Rupali Verma[2], Biswajeet Nayak[1], Suman Kumar Chakraborty[1], Shailendra Kumar Varshney,[3] Mayank Shrivastava[2], Prasana Kumar Sahoo[1*]

*[1]Quantum Materials and Device Research Lab, Materials Science Centre, Indian Institute of Technology Kharagpur, WB, 721302 India; [2]MSDLab, Department of Electronic Systems Engineering, Indian Institute of Science, Bangalore, 560012; [3]Electronics and Electrical Communication Engineering, Indian Institute of Technology Kharagpur, WB, 721302 India*

*prasana@matsc.iitkgp.ac.in

## Abstract

Heterogeneous integration typically relies on the co-integration of specialized devices for sensing, light generation, communication, and computation, resulting in complex interconnects, fabrication overhead, parasitic losses, and energy penalties. A longstanding goal of developing a monolithic, single reconfigurable device capable of selective carrier generation, transport, and recombination remains a challenge. 2D lateral heterostructures offer a promising route towards this goal by combining spatially distinct domains of different charge polarity whose electrostatic landscape can be dynamically engineered. Here, multifunctional bilayer $MoSe_2$–$WSe_2$ lateral heterojunction device architectures are demonstrated, integrating photodetection, reconfigurable EL, and CMOS logic operation within a single platform. As a photodetector, the device exhibits a responsivity of 35 $AW^{-1}$, a specific detectivity of $5\times10^{13}$ Jones, and a noise-equivalent power of $6.3\times10^{-18}$ $WHz^{-1/2}$, enabled by efficient charge separation and photoconductive gain. Under electrical carrier injection, the same heterointerface supports spatially programmable electroluminescence, in which electrostatic gating dynamically switches the radiative recombination zone between the $MoSe_2$ domain, the heterointerface, and the $WSe_2$ domain, enabling optical multiplexing and ternary optical logic functionality. Furthermore, the complementary transport characteristics of $MoSe_2$ and $WSe_2$ enable CMOS inverter operation with a switching frequency of 0.25 kHz. Remarkably, photodetection, light emission, and logic operation all emerge from a common electrostatic framework in which a gate-tunable depletion region controls carrier extraction, recombination, and transport. These results establish 2D lateral heterostructure as a multifunctional building block for monolithically integrated, energy-efficient electronics and optoelectronics systems that combine light sensing, generation, and information processing within a unified architecture.

## Introduction

The increasing demand for compact, energy-efficient, and multifunctional electronic systems on a single platform has intensified interest in heterogeneous integration as a strategy for next-generation electronics and optoelectronics[1–5]. In conventional technologies, functionalities such as sensing, communication, light generation, detection, memory, and computation are typically implemented using separate device architectures and interconnected at the circuit level[6]. While this approach enables functional specialization, it introduces fabrication complexity, parasitic interconnect loss, increased power consumption, and scaling limitations. Consequently, integrating heterogeneous functionalities within a unified material platform has emerged as a promising strategy for highly compact architecture and energy-efficient systems.

Two-dimensional (2D) materials are particularly attractive for next-generation heterogeneous integration owing to their atomic-scale thickness, tunable bandgap, high carrier mobility, better gate electrostatics, and dangling bond-free surfaces[7–9]. Significant progress in the synthesis and device engineering of transition-metal dichalcogenides (TMDs)[10–12] has enabled wafer-scale growth, high-performance p-type and n-type transistors with subthreshold slope approaching the thermodynamic limit (60 mV/dec), and integrated circuits such as CMOS inverters, operational amplifiers[13], ring oscillators[14], and microprocessors[15]. In addition, several heterogeneous integration strategies based on 2D materials have been explored, placing these materials on the technology roadmap of several semiconductor industries[16,17]. More broadly, 2D heterogeneous integration extends monolithic semiconductor integration to the atomic limit, enabling multifunctionality within a single platform. However, realizing multiple functionalities through a common physical mechanism within a single active device remains challenging.

Among the various approaches, vertical-stacked van der Waals (vdW) heterostructure[18–22] and monolithic three-dimensional (3D) integration[26–28] have emerged as the dominant strategies for heterogeneous integration using 2D materials. Vertical heterostructures combine distinct 2D layers to realize various functionalities, including photodetectors, light-emitting diodes (LEDs), logic, sensing, and memory, whereas monolithic 3D integration increases functional density through sequential device stacking. Although these approaches enable highly integrated systems, they often require complex fabrication processes and multiple active device units. In contrast, 2D lateral heterostructures (LHS) provide an alternative route for monolithic multifunctionality within a single device architecture. In particular, $MoSe_2$–$WSe_2$ LHS form atomically sharp in-plane p–n junctions, where $MoSe_2$ and $WSe_2$ domains typically behave as n-type and p-type 2D semiconductors, respectively[25]. Such a heterointerface generates a built-in electric field and a spatially separated gate-tunable depletion region that supports efficient modulation of carrier injection, in-plane transport, extraction, and recombination under different bias conditions.[26]. Early demonstrations of both lateral[27–31] and vertical[32] heterostructures established their potential for photodetection, exhibiting efficient charge separation and room-temperature responsivities ranging from sub-A/W[27–30] to several hundred A/W. However, the simultaneous realization of high responsivity and detectivity in photodetection, electrically reconfigurable light emission, electronic, and electro-optic logic functionality within a single LHS domain remains largely unexplored.

In most vdW heterostructure-based electroluminescence (EL) devices, the recombination zone is fixed by the material bandgap, the metal-semiconductor junction, or p-n junction, limiting dynamic control over light generation. In contrast, the gate-controlled depletion region in the $MoSe_2$–$WSe_2$ LHS enables spatially programmable recombination, allowing the emission spectra to be electrically switched among $MoSe_2$, heterointerface, and $WSe_2$ domains. More importantly, the same built-in electric field and gate-tunable electrostatic landscape simultaneously govern carrier separation during photodetection, radiative recombination during electroluminescence, and charge transport during logic operation, providing a unified physical mechanism for multifunctionalities within a single lateral device architecture.

Here, a multifunctional $MoSe_2$-$WSe_2$ LHS device is demonstrated that simultaneously exhibits; (i) FET operation, (ii) diode rectification, (iii) tunable high performance photodetection with a responsivity of 35 A/W with detectivity of ~$10^{13}$ Jones, (iv) spatially reconfigurable and electrically tunable EL enabling optical multiplexing, and (v) logic functionality including CMOS inverter with switching frequency up to 0.25 KHz, all within a single unified material platform. Furthermore, transient measurements provide insight into carrier recombination dynamics and their lifetime at the heterointerface. Collectively, these results establish $MoSe_2$-$WSe_2$ LHS as a multifunctional unit-device building block for future atomically thin heterogeneous integration platforms combining sensing, light generation, and information processing within a unified architecture.

**Results & Discussion**

Bilayer $MoSe_2$-$WSe_2$ LHSs were directly grown on Si($n^{++}$)/$SiO_2$ substrates using chemical vapor deposition via one-pot edge-epitaxy[33], forming alternating $MoSe_2$ and $WSe_2$ domains with atomically sharp interfaces (Fig. 1a). Atomic-resolution high-angle annular dark-field scanning transmission electron microscopy (HAADF-STEM) imaging confirms the formation of sharp 1D lateral interfaces between the $MoSe_2$ and $WSe_2$ domains (Fig 1b). The distinct Z-contrast clearly distinguishes the darker $MoSe_2$ and brighter $WSe_2$ regions, arising from differences in their atomic numbers. Several devices were fabricated using the 2D LHSs as the active channel, enabling systematic studies of charge transport, photodetection, EL, and logic operation.

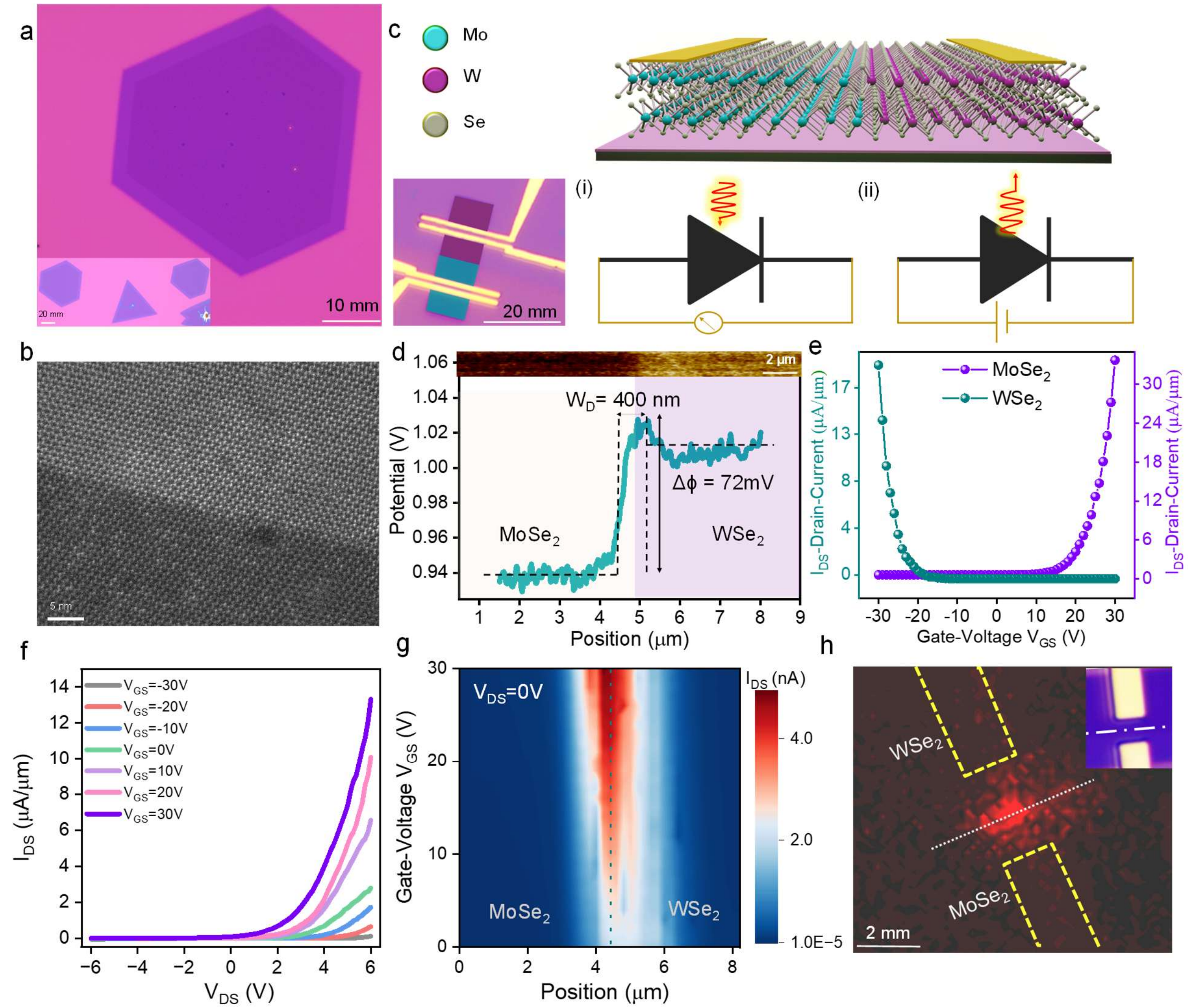

***Figure 1| Characteristics of Bilayer $MoSe_2$–$WSe_2$ lateral heterostructure (LHS)***. ***a***, *Optical microscope image of a representative bilayer $MoSe_2$–$WSe_2$ lateral heterostructure. Insets show a low-magnification view of the as-grown LHs on a single substrate. b, High-resolution transmission electron microscopy image of the heterointerface, confirming the atomically seamless epitaxial junction.* ***c***, *Atomic model of the $MoSe_2$–$WSe_2$ LHS device and the corresponding optical microscopic image (bottom left. Schematics illustrate the operating principles of the fabricated device (false color) under (i) photodetection and (ii) electroluminescence (bottom right).* ***d***, *Kelvin probe force microscopy mapping of the contact potential profile measured across the device-heterointerface, revealing a depletion width of ~400 nm and a built-in potential difference of ~72 mV.* ***e***, *Transfer characteristics of the individual $MoSe_2$ and $WSe_2$ regions, demonstrating complementary n-type and p-type transport, respectively.* ***f,*** *Gate-dependent output characteristics $I_{DS}$–$V_{DS}$ of the LHS, showing pronounced rectifying behavior.* ***g***, *short-circuit photocurrent map as a function of position and gate voltage, revealing strong photocurrent generation localized at the heterointerface.* ***h***, *Spatial electroluminescence image acquired under electrical injection, showing localized light emission from the heterojunction region where electrons and holes recombine radiatively.*

Owing to type-II band alignment and complementary carrier polarities of these CVD-grown $MoSe_2$ and $WSe_2$ bilayers, the 2D LHS forms a built-in electric field and depletion region at the atomically sharp p–n heterojunction. Under optical excitation, this built-in field separates and extracts photogenerated carriers, enabling photodetection (*Fig. 1c (i)*). Conversely, under electrical injection, electrons and holes are driven towards the depletion region, where they recombine radiatively, resulting in EL (*Fig. 1c(ii)*). To probe the junction electrostatics, Kelvin probe force microscopy was used to map the surface potential across the lateral $MoSe_2$–$WSe_2$ interface, revealing a built-in potential difference of ~72mV and a depletion width of ~400 nm (*Fig. 1d*). It reflects the interfacial band bending arising from the type-II band alignment, which facilitates efficient charge separation. The space charge region plays a central role in the multifunctional device operation by governing carrier generation, extraction, injection, and recombination.

The transfer characteristics ($I_{DS}$-$V_{GS}$) of the FETs fabricated on the individual $WSe_2$ and $MoSe_2$ domains exhibit complementary p-type and n-type behavior, respectively (*Fig. 1e; S1|d-e*). Both devices show typical $I_{on/off}$ ratios approaching ~$10^8$, indicative of excellent gate controllability and complementary carrier polarity, which are essential for PN junction formation and logic functionality. The *IV* characteristics measured across the heterojunction show diode-like rectification, confirming the formation of a P-N junction diode. Under positive $V_{DS}$ (i.e., forward bias), $I_{DS}$ increases exponentially, whereas under reverse bias, $I_{DS}$ is significantly suppressed (primarily due to minority-carrier transport), resulting in strong rectification. The reproducibility of the rectifying behavior is further confirmed by statistical measurements across multiple devices (*Fig. S1|f)*

To directly visualize the active carrier-separation region, spatial photocurrent mapping was performed under short-circuit conditions ($V_{DS}=0V$) across the PN junction device using a *532nm* laser excitation *(Fig. 1g)*. The short-circuit photocurrent ($I_{SC}$ ) provides a direct measure of the intrinsic junction photoresponse, governed primarily by the built-in electric field and interfacial band alignment. The applied vertical field modulates the carrier concentration according to the overdrive voltage $V_{OV}=V_{GS}-V_T$. Maximum photocurrents are localized at the heterointerface and coincide with the depletion region identified by KPFM. Upon illumination, photogenerated electron-hole pairs are extracted by the built-in field, with electrons drifting towards the $MoSe_2$ and holes towards the $WSe_2$ regime. This direct visualization of the photocurrent profile and the surface potential map establishes the heterointerface as the active carrier-extraction region. The same heterointerface also functions as an active carrier recombination under forward bias. EL imaging reveals pronounced localized light emission from the junction region *(Fig. 1h)*, confirming efficient radiative recombination of injected electrons and holes within the depletion region. The spatial EL emission confirms that the junction serves as an active recombination zone, analogous to its role as an active carrier-extraction region during photodetection. The coexistence at the same heterointerface highlights the multifunctionality of the lateral PN junction.

## Gate modulated photodetection

Spatial photocurrent mapping with varying $V_{DS}$ is used to map the $V_{GS}$-dependent band bending and probe the role of electrostatic doping in photocarrier generation and extraction (Fig. 2a–c). Changing $V_{GS}$ modulates the carrier concentration in the $MoSe_2$ and $WSe_2$ domains, thereby altering the depletion region, the junction field, and the carrier collection efficiency. At $V_{GS}=0V$, the photocurrent is strongly localized at the junction, where the in-built field is maximum, leading to pronounced carrier extraction (*Fig. 2b)*. Under negative $V_{GS}$ *(-30V)*, more hole-accumulation occurs in the $WSe_2$ domain, whereas the depletion of electrons in the $MoSe_2$ expands the space-charge region towards the n-type side. The corresponding overall electric-field redistribution causes the photocurrent maxima to shift towards the $MoSe_2$ side (*Fig. 2a)*. Conversely, positive $V_{GS}$ increases the electron accumulation in the $MoSe_2$. In contrast, it depletes the hole concentration in the $WSe_2$ region, resulting in a shift of the depletion region towards the $WSe_2$ side (*Fig. 2c).* Hence, a corresponding increase in the photocurrent shifts is observed towards the $WSe_2$ domain. The gate-dependent displacement of the photocurrent maximum directly visualizes the electrostatic control of the depletion region and confirms its central role in carrier extraction.

To quantify the photoresponse, the heterointerface was selectively illuminated while measuring the photocurrent under different drain and gate biases (*Figs. 2d, S2a)*. The enhancement of net $I_{DS}$ compared to the dark current confirms efficient photocarrier generation and extraction at the active interface region (*Fig. S2a). The s*tatistical distribution shows device-to-device variation in photocurrent, measured at the same power, confirming the reproducibility of the photoresponse across multiple devices (*Fig. 2d*). The photocurrent exhibits strong dependence on both optical power (100 nW to 32 μW) and $V_{GS}$ (*Figs. S2-4)*, reflecting the interplay between carrier generation, space-charge modulation, carrier extraction, and photoconductive gain. The power-dependent $I_{PC}$ at $V_{GS}$=0V shows the strong localization of current at the heterointerface, further confirming the role of the built-in electric field at the lateral p–n junction (Fig. S2 b, c). Moreover, under reverse bias, the strengthened junction field and widened depletion region enhance photocarrier extraction while suppressing recombination losses, yielding a higher photoresponse than in forward bias, even in the absence of VGS. Irrespective of gate polarity (VGS=±30V), forward-bias operation yields a larger net current than reverse bias (Figs. S3,4) due to more efficient carrier injection across the junction.

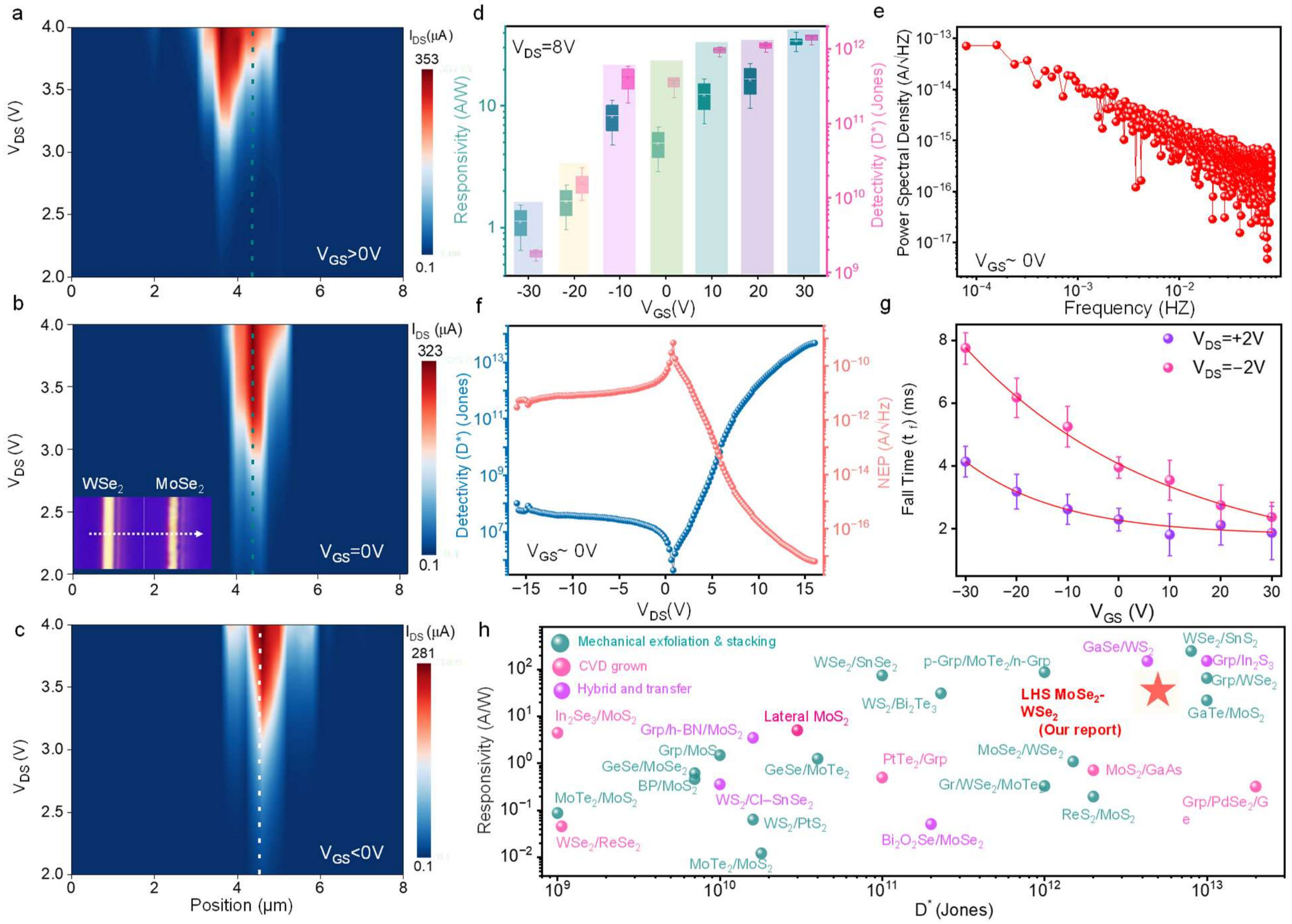

***Figure 2| Gate-tunable photodetection characteristics of the bilayer $MoSe_2$–$WSe_2$ lateral heterojunction**. Spatial photocurrent $I_{PC}$ maps measured as a function of drain bias ($V_{DS}$) at different $V_{GS}$ of (a) =-30 V, (b) 0 V, and (c) +30 V, revealing that the $I_{PC}$ is strongly localized at the heterointerface, where the built-in electric field efficiently separates photogenerated electron–hole pairs. As the $V_{GS}$ varied from negative to positive values, the $I_{PC}$ distribution progressively shifted across the junction, reflecting gate-induced modulation of the band alignment, depletion region, and carrier collection profile. The inset in b shows the device geometry and scanning direction used for $I_{PC}$ mapping. d, Photoresponsivity (R) and specific detectivity ($D_{spec}$*) as a function of $V_{GS}$ under 8 μW illumination at $V_{DS}$=15V, both showing substantial enhancement with increasing positive $V_{GS}$. e. Frequency-dependent noise-current power spectral density, showing dominant low-frequency 1/f noise. f, Specific detectivity (D*) extracted from the measured noise data as a function of $V_{DS}$, reaching values approaching $10^{14}$Jones and corresponding noise-equivalent power (NEP) g. The fall time $t_f$ as a function of $V_{GS}$ under $V_{DS}$ = +2V, and j. $V_{DS}$ = -2V, demonstrating progressively faster response dynamics at higher $V_{GS}$ as a result of enhanced carrier transport and extraction across the heterojunction. The reduction is more pronounced under reverse bias, resulting in shorter response times due to more efficient carrier sweep-out by the junction electric field. h. Comparative study of photodetection performance of the $MoSe_2$-$WSe_2$ lateral heterostructure based devices with reported photodetectors in terms of responsivity and detectivity[34–60].*

Similarly, under $V_{GS}<0V$, the $I_{PC}$ increases with increasing optical power exposure along with $V_{DS}$ (*Fig. S3*). With $V_{GS}>0V$ under forward bias, the reduced junction barrier upon increasing illumination substantially increases $I_{PC}$ (*Fig. S4). In contrast, under* reverse bias, although the enlarged depletion region efficiently separates the photogenerated carriers through drift transport, the photoresponse remains limited by the carrier generation rate. The transient photodetection analysis at $V_{DS}=0V$ reveals that the increase in $I_{PC}$ with increasing exposure power is driven by a large number of photogenerated carriers, which are promptly separated by the lateral field and collected at the electrode. The power-dependent photocurrent initially increases nearly linearly with optical power, then approaches saturation above 52 mW (*Fig. S2d*). This behavior originates from the interplay between carrier generation, recombination, and field screening at the heterointerface. At high carrier densities, accumulated photocarriers partially screen the built-in electric field, reducing the efficiency of carrier separation and limiting further $I_{PC}$ enhancement.

The wavelength-dependent photoresponse was investigated using red (625 nm), green (532 nm), and blue (470 nm) illumination at different $V_{GS}$ (*Figs. S5)*. The strongest response is obtained under red excitation (with the lowest optical power among the three), which lies closest to the near-band-edge absorption in $MoSe_2$ and $WSe_2$, where excitonic transitions dominate[61]. In contrast, the excited carriers generated by higher-energy green and blue photons undergo rapid carrier thermalization and enhanced recombination, resulting in reduced $I_{PC}$[62]. Gate and drain biases further modulate the $I_{PC}$ by altering the depletion region and built-in field, with reverse bias consistently yielding the strongest response due to more efficient carrier separation and collection.

Using the extracted $I_{PC}$, the photoresponsivity (R) and specific detectivity (D*) were calculated using $D_{spec}^{*} = R\sqrt{A} \,/\, \sqrt{2qi_d}$ where A denotes the channel area ($W_{CH}*L_{CH}$), and $i_d$ is the dark current. The device exhibits a responsivity of *35A/W* at an optical power of *8mW* and D* approaching ~$10^{12}$ *Jones* for $W_{CH}$ *of 3μm* and $L_{CH}$ *of 3μm* (*Fig. 2d)*, indicating efficient photocarrier generation and collection at the heterointerface. Detectivity increases with drain bias owing to the combined effect of efficient extraction of photogenerated carriers by the enhanced lateral electric field and bias-dependent photoconductive amplification. Interestingly, the highest responsivity and detectivity are achieved under forward bias than in reverse bias, unlike conventional photodiodes. This behavior arises from injection-assisted photoconductive gain, where the $I_{PC}$ is governed not only by carrier separation at the junction but also by the large disparity between carrier transit and recombination times. Consequently, the enhanced photoresponse originates from the combined effect of efficient charge separation and photoconductive gain, as confirmed by the transient photodetection analysis discussed below.

To evaluate the sensitivity limit of the heterojunction photodetector, the current noise spectral density was measured across a broad frequency range (*Fig. 2e*), from which the $D^*_{spec}$ and noise-equivalent power ($NEP = \frac{i_{noise}}{R}$) were calculated as a function of $V_{DS}$. The device exhibits a maximum detectivity of $\sim 5\times10^{13}$ Jones with a minimum NEP of $\sim 6.3\times10^{-18}\ W\ Hz^{-1/2}$ (*Figs. 2f*), highlighting its capability for ultra-low-light detection. As $V_{DS}$ increases, the $I_{PC}$ increases significantly compared to the noise current, thereby improving the overall signal-to-noise ratio and yielding superior detectivity. The simultaneous realization of high responsivity, ultra-low NEP, and detectivity approaching $\sim 10^{14}$ Jones highlights the effectiveness of combining built-in carrier separation with gain-assisted transport in a single $MoSe_2$–$WSe_2$ LHS. Full-area illumination measurements further confirm that the heterointerface dominates the photoresponse (*Fig. S6*), while comparison with state-of-the-art 2D photodetectors demonstrates competitive overall performance with key figure of merit responsivity and detectivity[34–60] *(Fig. 2h).*

The transient $I_{PC}$ measurements were performed at $V_{DS}=\pm 2V$ to evaluate the dynamic response of LHS photodetectors (*Fig. 2g)*. The device exhibits millisecond-scale rise and fall times ($t_f$), whereas the carrier transit time ($\tau_{transit} = L/\mu.E$; $E = \frac{v_d}{L}$), across the channel, is estimated to be in the sub-nanosecond regime. The six-order-of-magnitude difference between transit and response times indicates extremely efficient carrier extraction and substantial photoconductive gain. Accordingly, the gain, estimated as $G\approx\tau_f/\tau_{transit}$, *approaches* $\sim 10^6$, implying that a single photogenerated carrier can contribute multiple times to the measured current before recombination[63]. Furthermore, as $V_{GS}$ changes gradually from negative to positive values, the fall time decreases exponentially (*Fig. S8a-c)*, owing to the electrostatic modulation of the carrier density in the 2D channel, according to the equation $n(V_{GS}) = (C_{ox}/q).(V_{GS}-V_T)$ [64].

The trap-assisted recombination lifetime follows $t_{rec}\propto 1/n$, indicating that an increase in carrier density accelerates shallow trap-assisted recombination, leading to faster decay of photocurrent and hence the fall time[65]. The effect is more pronounced under reverse bias ($V_{DS}=-2V$), where the channel is more strongly depleted and the initial carrier density is lower, resulting in longer recombination lifetime and greater sensitivity of $t_f$ to $V_{GS}$ (Fig. 2g). In contrast, carrier injection under forward bias ($V_{DS}$=+2$V$) increases the background carrier density, thereby reducing the relative modulation of the recombination rate (Fig. 2g). Overall, the negligible dependence of fall time on $V_{DS}$, compared with its strong dependence on $V_{GS}$, indicates that the transient response is dominated by recombination-limited. The coexistence of nanosecond transit times and millisecond fall time indicates the device operates in a photoconductive regime, dominated by trap-assisted recombination with electrostatic transport prevailing over drift-based extraction. Such gate-tunable carrier dynamics provide a means to engineer the gain–speed trade-off while simultaneously enhancing responsivity, detectivity, and photoconductive gain through electrostatic control of the interfacial depletion region.

## Electroluminescence (EL)

The same LHS device that drives photo-carrier separation also serves as a gate-reconfigurable radiative electroluminescent junction. Under forward bias, electrons and holes, injected from the $MoSe_2$ and $WSe_2$ domains, respectively, recombine to produce EL, while electrostatic gating dynamically controls the injection and recombination pathways. Consequently, unlike conventional 2D LEDs with fixed emission characteristics determined by the material bandgap or junction geometry, the 2D LHS supports electrically programmable emission (*Figs. 3a-e*). At $V_{GS}$*=0V,* the EL originates predominantly at the $MoSe_2$-$WSe_2$ heterointerface, where carrier recombination is maximized (*Fig. 3c*). Applying a negative $V_{GS}$ *(≈-15, -30V)* depletes the electron concentration in the $MoSe_2$ side and increases the hole concentration in the $WSe_2$ region, shifting the dominant recombination zone and EL emission into the $MoSe_2$ region (*Figs. 3a, b*). Conversely, at $V_{GS}$*>0V*, electron accumulation in $MoSe_2$ increases, whereas hole concentration decreases in the $WSe_2$ side, shifting the depletion zone to the $WSe_2$ side, where the EL emission is predominantly observed (Figs. 3d, e). The spatial EL imaging and the corresponding spectra (Figs. 3f-h) further confirm this gate-controlled evolution of the recombination zone and emission wavelength. Similar behavior is observed across multiple devices, with representative electrical characteristics, EL spectra, and band diagrams provided in the *Figs. S9 and S10*. In addition, constant current injection over an extended period for all investigated $V_{GS}$ (*Fig. S11)* demonstrates stable EL operation.

Interestingly, EL is also observed under reverse-bias ($V_{GS}$*<0V*) operation, beyond the junction breakdown regime (Fig. *S12*). In this regime, the high electric field generates additional electron-hole pairs via impact ionization. Under $V_{GS}$<0V, the hot-carrier relaxation and subsequent radiative recombination occur predominantly within $MoSe_2$, resulting in intense reverse-bias emission. Moreover, the EL intensity increases monotonically with injection current due to efficient radiative recombination (*Fig. S13)*.

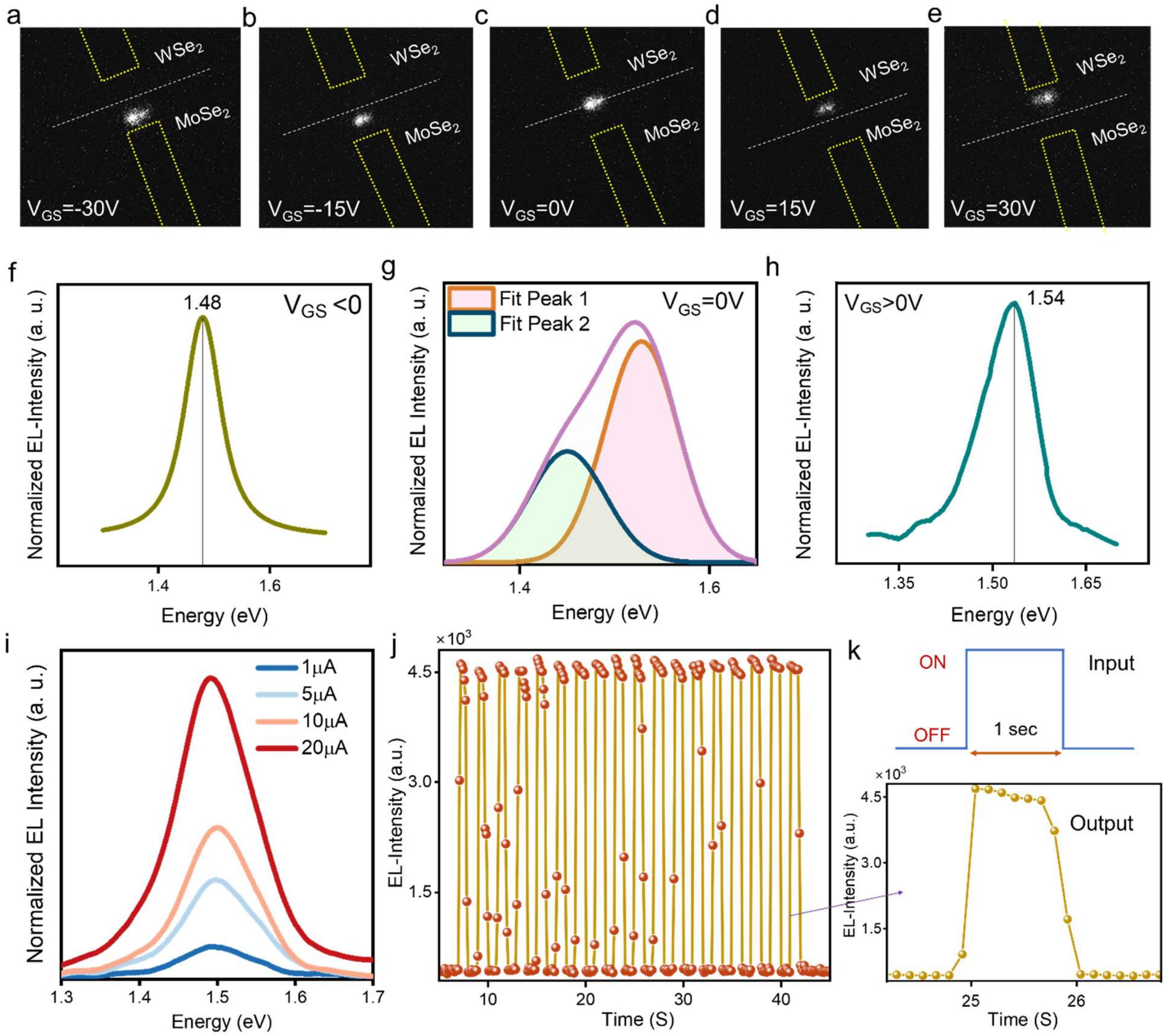

***Figure 3 | Gate-tunable electroluminescence characteristics of the bilayer $MoSe_2$–$WSe_2$ lateral heterojunction.*** *(**a–e**) Electroluminescence (EL) maps acquired at $V_{GS}$=-30, -15, 0, +15, and +30 V, revealing a systematic shift of the emission zone across the heterojunction. The EL emission is predominantly localized on the $MoSe_2$ side under negative $V_{GS}$, moves towards the heterointerface at $V_{GS}$=0 V, and progressively shifts towards the $WSe_2$ side under positive $V_{GS}$, demonstrating electrostatic control of the carrier recombination region; and **f–h,** the corresponding EL spectra acquired from the dominant emission regions at different $V_{GS}$. **f**, At $V_{GS}$ <0 V, the emission energy corresponds to the $MoSe_2$; **g**, At $V_{GS}$ =0 V, the emission is localized at the heterointerface and shows contributions from both constituent materials; and **h**, At $V_{GS}$ >0 V, the emission shifts to the $WSe_2$ domain and is dominated by the characteristic of $WSe_2$ emission peak. **i**, Gate-dependent EL spectra showing a pronounced enhancement in emission intensity with increasing injection current. **j**, Time-resolved EL intensity measured under periodic electrical excitation, demonstrating stable and reproducible light emission over multiple switching cycles. **k**, Schematic illustration of the electrical pulse sequence used for dynamic EL measurements. **l**, enlarged view of a single switching cycle, confirming the rapid turn-on and turn-off behavior of the 2D heterojunction.*

To investigate the dynamic EL response, time-resolved EL measurements were performed under periodically pulsed forward-bias excitation at room temperature. The EL intensity exhibits reproducible emission bursts synchronized with the electrical pulses (ν=1Hz), demonstrating deterministic electrical control of light generation (*Figs. 2 j,k)*. Upon pulse application, the EL rapidly rises to a steady-state intensity of ~$4.5x10^3$ *a.u.* and remains stable throughout the high-voltage (ON-state) period, indicating a balance between carrier injection and recombination. Notably, the EL extinguished immediately after the pulse is removed (OFF-state), with no measurable decay between the successive cycles. This behavior confirms that the EL is primarily governed by direct radiative recombination of electrically injected carriers rather than a trap-assisted process. The sharp EL switching reveals that the recombination dynamics are considerably faster than the applied timescale between electrical excitation. Similar switching behavior is observed across pulse periods ranging from 10 ms to 1 s, although the intrinsic recombination lifetime remains below the temporal resolution limitation of the measurement setup (Supplementary video 1-2). The results demonstrate effective reversible switching between the emissive 'ON' state and the non-emissive 'OFF' state, thereby representing the electrically controlled optical switching behavior.

The gate-tunable EL can be directly correlated with the operation of an optical multiplexer (MUX) and multivalued logic functionality. By electrostatically shifting the recombination zone, a single analog gate input ($V_{GS}$) selectively routes EL output to the $MoSe_2$ ($D_{-1}$, $V_{GS}<0$), the lateral interface ($D_0$, $V_{GS}=0$), or the $WSe_2$ ($D_{+1}$, $V_{GS}>0$) (*Figs. 4a,b*). This behavior naturally defines a ternary logic scheme with logic states *-1, 0,* and *+1* encoded by emission from $MoSe_2$, lateral interface, and $WSe_2$, respectively (*Fig. 4c*). The corresponding ternary truth table and Karnaugh map validate the operation of a one-input and three-output optical selector, equivalent to a signed ternary multiplexer (*Fig. 4d*).

Unlike conventional multivalued photonic circuits[66,67] that require multiple active components, the present LHS directly encodes spectral and spatial information within a single device architecture, inherently expanding the logic capacity to a ternary logic system. Such multiple discrete logic states can enable high data processing efficiency and reduced complexity.[68] Such electrostatically programmable light emission provides a compact platform for on-chip optical routing, low-power optoelectronic multiplexing, and area-efficient future multivalued logic based on 2D LHS.

a

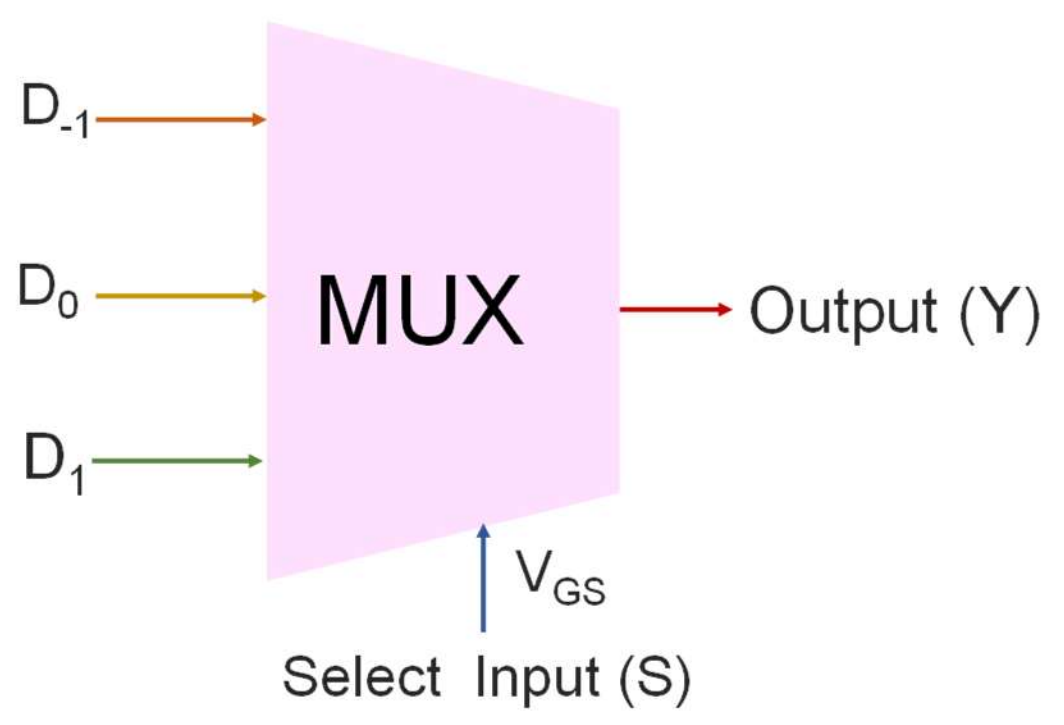

b

| $V_{GS}$ (s) | EL Region | I (Emission Peak) | Ternary Logic |
|---|---|---|---|
| < 0 V | $MoSe_2$ ($D_{-1}$) | ~1.48eV | –1 |
| = 0 V | Interface ($MoSe_2$–$WSe_2$) ($D_0$) | ~1.4-1.6eV | 0 |
| > 0 V | $WSe_2$ ($D_1$) | ~1.54eV | +1 |

c

| S | $D_{-1}$ | $D_0$ | $D_{+1}$ | Y | |
|---|---|---|---|---|---|
| –1 | 0 | X | X | 0 | $D_{-1}$ |
| –1 | 1 | X | X | 1 | |
| 0 | X | 0 | X | 0 | $D_0$ |
| 0 | X | 1 | X | 1 | |
| +1 | X | X | 0 | 0 | $D_1$ |
| +1 | X | X | 1 | 1 | |

d

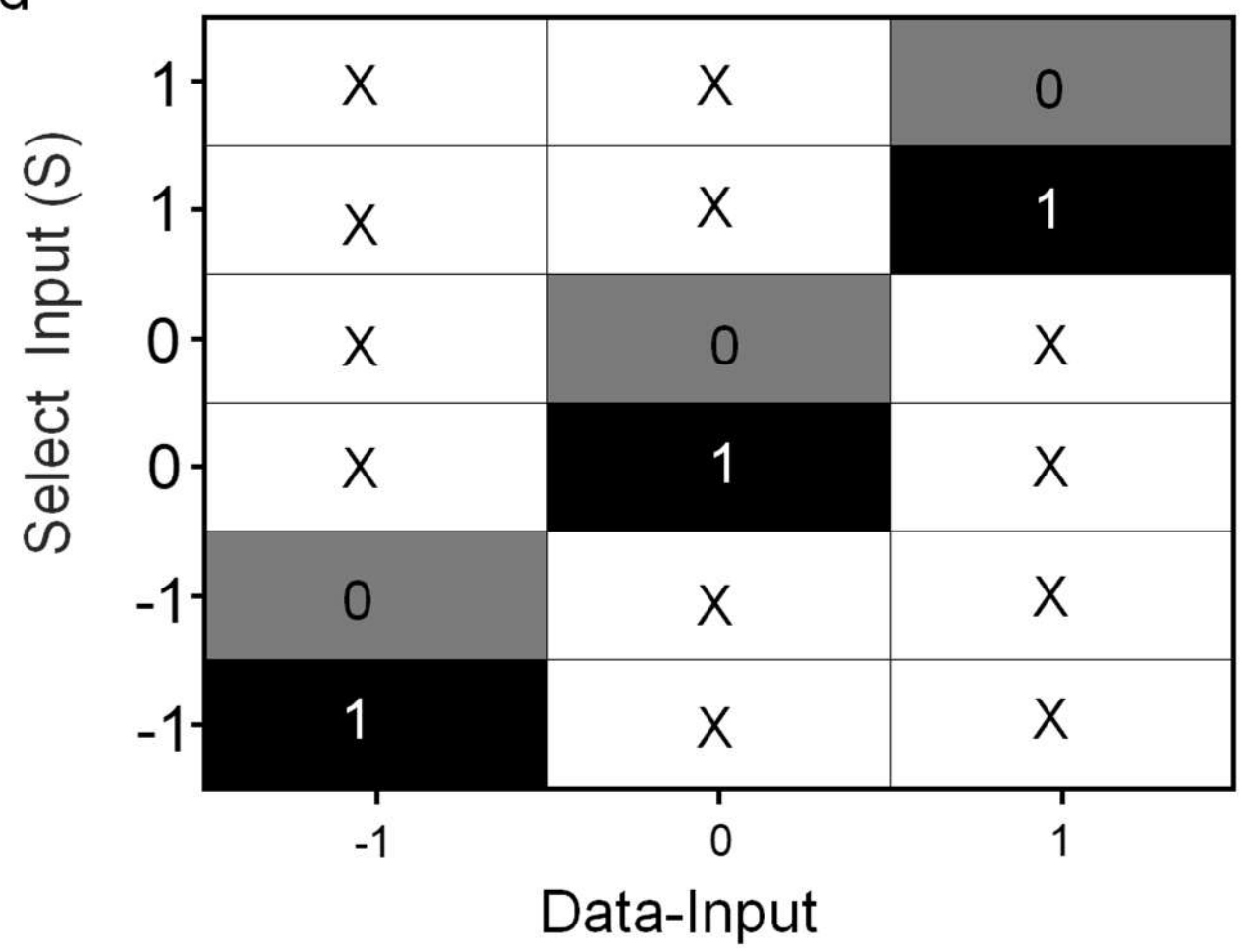

***Figure 4 | Electroluminescent multiplexer operation based on the bilayer $MoSe_2$–$WSe_2$ lateral heterojunction. a,*** *Schematic illustration of the multiplexer, where the $V_{GS}$ serves as the select input and controls the output emission channel.* ***b,*** *Correspondence between $V_{GS}$, EL emission region, emission wavelength, and ternary logic states. Negative, zero, or positive $V_{GS}$ select emission from the $MoSe_2$ domain $D_{-1}$, heterointerface $D_0$, or $WSe_2$ domain $D_{+1}$, respectively.* ***c,*** *Truth table describing the ternary multiplexer operation.* ***d,*** *Logic matrix illustrating the gate-controlled selection of the three electroluminescent output states.*

## CMOS Inverter

Beyond photodetection, electroluminescence, and multiplexing, practical heterogeneous integration also requires on-chip computation. Owing to the complementary carrier polarity of $MoSe_2$ and $WSe_2$ domains within LHS, it can be configured to perform CMOS logic operations. The p-type and n-type of $WSe_2$ and $MoSe_2$ domains were configured into a four-terminal CMOS inverter geometry (*Fig. 5a)*. The source contacts of $WSe_2$ (electrode 1) and $MoSe_2$ (electrode 4) are connected to $V_{DD}$ and ground (*GND*), respectively, while their drain contact electrodes (2 and 3) were electrically sorted to form the output node ($V_{OUT}$) (*Figs. 5a, b*). The common $V_{GS}$ serves as the input signal, controlling both the transistors simultaneously. The voltage transfer characteristics (VTC) show the typical inverter characteristics (*Fig. 5c)*. At $V_{GS}<0V$, the $WSe_2$ as PMOS is ON, while the $MoSe_2$ NMOS remains OFF, pulling the $V_{OUT}$ close to $V_{DD}$ and producing a logic 'HIGH' state. As $V_{GS}$ increases, PMOS gradually turns 'OFF', while NMOS turns 'ON', resulting in a sharp transition of $V_{OUT}$ toward *GND* and generating a logic 'LOW' state (*Fig. 5c)*. The observed switching transition confirms effective complementary operation enabled by the intrinsic p-type and n-type characteristics of the $WSe_2$ and $MoSe_2$ domains, respectively.

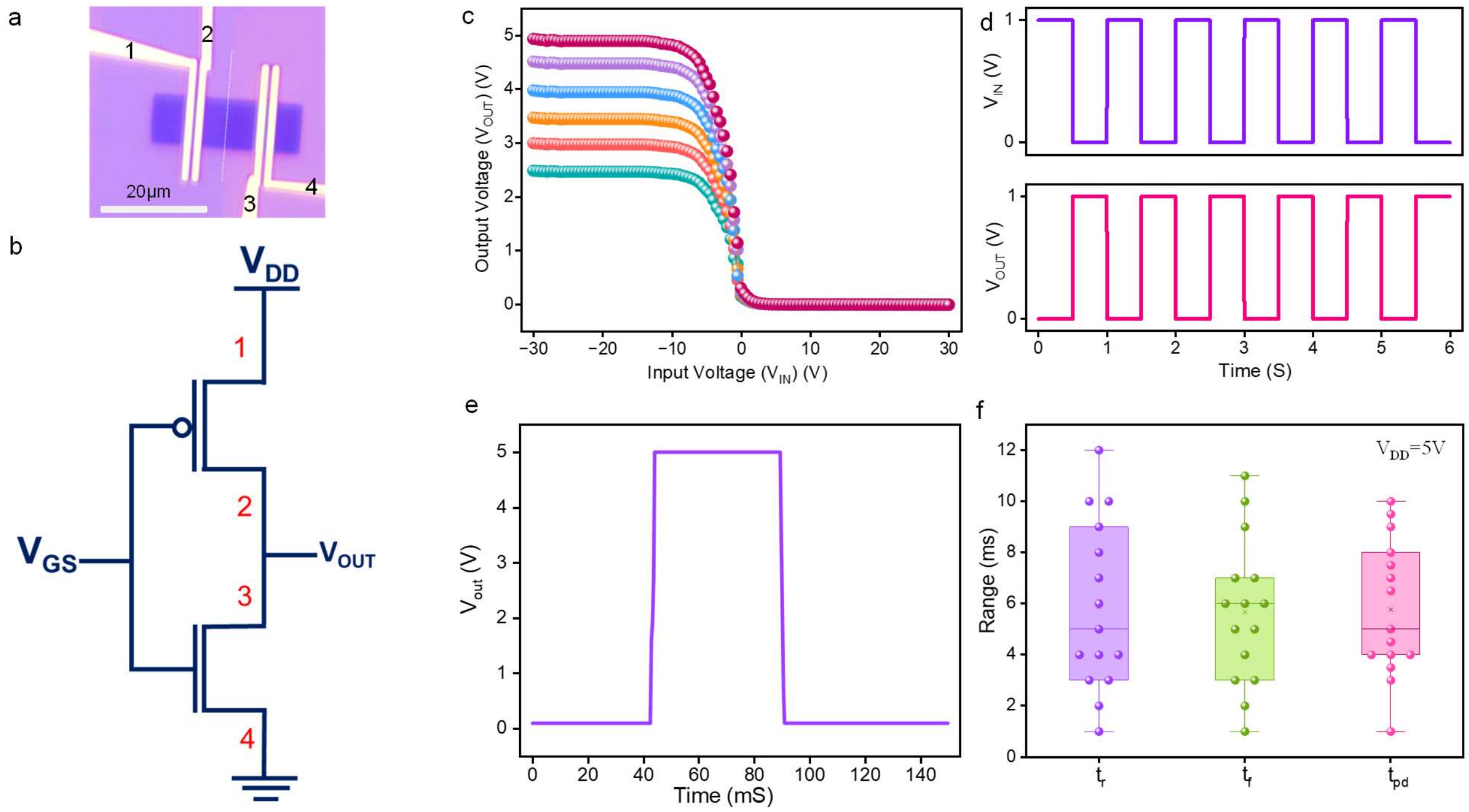

***Figure 5 | Complementary inverter based on the bilayer $MoSe_2$–$WSe_2$ lateral heterojunction.*** *a, Optical microscope image of the fabricated inverter device. Scale bar, 20 μm. b, Circuit schematic of the complementary $MoSe_2$–$WSe_2$ inverter showing the contact configuration and measurement geometry. c, Voltage transfer characteristics ($V_{OUT}$–$V_{IN}$) measured at different supply voltages ($V_{DD}$), exhibiting a clear transition between the logic HIGH and logic LOW states. d, Transient inverter operation under periodic square-wave input, showing the complementary relationship between the input and output signals. e, Representative time-resolved output voltage waveform demonstrating stable and repeatable switching between the HIGH and LOW logic states at $V_{DD}$=5V. f, Statistical distribution of the extracted inverter performance parameters obtained from multiple switching cycles, confirming the reproducibility and reliability of the device operation*

The transient switching was further verified by applying a square-wave input signal to $V_{GS}$. The corresponding output waveform exhibits a clear input-signal (0→1) inversion, confirming CMOS inverter functionality (*Fig. 5d)*. The transient characteristics under a supply voltage $V_{DD}$=5V of the inverter switches between a logic "HIGH" state of 5V and a logic "LOW" state of 0.2V, yielding a large voltage swing of 4.8V (*Fig. 5e, S14-15)*. The finite rise and fall edges arise from the charging and discharging dynamics of the $V_{OUT}$ node through the effective channel resistance of the complementary transistors (*Fig. 5e* and additional *Fig. S15)*. The extracted rise ($t_r$) and fall time ($t_f$) are ~ 1-3 ms and ~1-2 ms, respectively (*Fig. 5e* and additional *Fig. S15),* corresponding to a propagation delay ($t_d$) of ~1-2 ms and operating frequency ($t_d^{-1}$) of ~0.25 kHz. This behavior is consistent with RC-limited switching dynamics of the inverter. The statistical analysis across multiple devices confirms the reproducibility and robustness of the 2D LHS CMOS operation (*Fig. 5f)*.

Together with spatially and spectrally tunable photodetection, electrically tunable EL optical multiplexing functionalities, and CMOS logic operation, establishes the $MoSe_2$-$WSe_2$ LHS as a multifunctional building block capable of integrating sensing, emission, and logic within a single atomically thin material platform, paving the way for future all-in-plane heterogeneous integration.

**Conclusion**

An atomically thin $MoSe_2$-$WSe_2$ bilayer LHS is shown to integrate photodetection, electrically programmable electroluminescence, optical multiplexing, and CMOS logic operation within a single device architecture. Structural and electrostatic characterization confirms an atomically sharp lateral P-N junction with a rectification ratio of $\sim 10^6$, a built-in potential barrier of ~ 72mV, and a gate-tunable depletion region. Spatial photocurrent mapping directly visualizes the modulation of the junction electrostatics, revealing that carrier generation, separation, and extraction are governed by the junction electric field. Consequently, the device exhibits responsivity up to *~35A/W* and detectivity approaching $\sim 10^{13}$ Jones. The same device also supports electrically programmable electroluminescence, where gate-controlled recombination enables both spatial and spectral tuning of the emission. This functionality further enables optical multiplexing (MUX) and ternary optical logic functionality. Time-resolved measurement reveals deterministic electrical control of emission and optical switching behavior. Furthermore, the complementary carrier polarities of $MoSe_2$ and $WSe_2$ enable CMOS inverter operation with a large voltage swing of ~ *4.8V* and a switching frequency of ~ *0.25 kHz*.

Collectively, these results establish a single lateral heterojunction as a unified platform in which carrier separation, recombination, and logic operation are governed by a common electrostatic framework. The ability to integrate sensing, emission, and computation into a single atomically thin device highlights the potential of 2D lateral heterostructures as a fundamental building block for future heterogeneous integration, compact optoelectronic systems, energy-efficient, and multifunctional 2D integrated circuits.

## Methods

### Direct Synthesis of 2D Lateral Multijunction Heterostructures

Bilayer 2D LHS were synthesized directly via a robust, scalable, water-assisted chemical vapor deposition (CVD) method. Bulk $MoSe_2$ (99.99%) and $WSe_2$ (99.99%) precursors were used as sources. $Si^{++}/SiO_2$ (285 nm) substrates, cleaned sequentially in acetone, isopropanol, and deionized water, were placed within a 1-inch quartz tube furnace, ~5 cm downstream from the precursors. A mixture of source precursors ($MoSe_2$:$WSe_2$: 60:40 wt%) at zone 1 was heated to 1050 °C, while the substrates were maintained at 750-800 °C. A 200 sccm flow of ultrahigh-purity $N_2$ (99.999%) initially purged the tube for 30 min, then the temperature of Zone 1 was raised. Upon reaching 1000 °C, $N_2$ diverted through a water bubbler at a flow rate of 200 sccm was introduced to oxidize the precursors. This oxidation generated volatile transition metal oxides and chalcogen species, initiating $MoSe_2$ growth. Abruptly switching the carrier gas to Ar+$H_2$(5%) suppressed further growth of the $MoSe_2$ domain. Meanwhile, $WSe_2$ with high-index tungsten-oxide-related precursors, formed during the oxidation stage, converted to lower sub-oxides. These preferentially deposited at $MoSe_2$ edges, enabling lateral $MoSe_2$-$WSe_2$ growth. Sequential carrier gas switching thus enabled the direct fabrication of multijunction heterostructure formation via self-aligned lateral epitaxy.

### Structural and Optical characterization

The Raman and photoluminescence (PL) Spectra were acquired by a LABRam Horiba Spectrometer with a 532nm laser. A laser with 0.8μW of power was focused through a 100x objective, and the emission spectra were collected using gratings with 1800 grooves/mm and 600 grooves/mm. For an atomic-resolution image, as-grown 2D LHS were transferred onto TEM grids by a PMMA-assisted wet-transfer process. Aberration-corrected high-angle annular dark-field scanning transmission electron microscopy (HAADF-STEM) was performed using a JEOL JEM-ARM300F2 microscope equipped with a cold-field emission gun operated at an accelerating voltage of 300 kV. HAADF-STEM images were taken using a JEOL HAADF detector, with a probe size of 8c and a scan speed of 20 μs/ pixel.

### Device Fabrication

The as-grown 2D LHS flakes were first patterned to define the device active regions and etched to selectively access individual $MoSe_2$ and $WSe_2$ domains. First, PMMA 495 A4/950 A2 positive-resist was spin-coated onto the substrate and patterned using electron-beam lithography, then developed in an MIBK: IPA (1:3) solution. The exposed regions were removed using oxygen plasma etching in a Reactive Ion Etching chamber ($O_2$ flow: 20sccm, RF power: 20W, and Time: 40s). After etching, the remaining resist was removed by soaking the substrate in acetone for 24 hours. Subsequently, electron-beam lithography steps were employed to define the active channel regions for individual $MoSe_2$ and $WSe_2$ domains, for asymmetric metal-contact patterns. Ni/Au (20/50 nm) contacts for $MoSe_2$ and Pd/Au (10/60 nm) for $WSe_2$ were deposited by e-beam evaporation (Tecport) at a pressure of $2e^{-6}$ torr, followed by lift-off in acetone. Finally, fabricated devices were vacuum-annealed at $1e^{-6}$ torr and 200 $^{o}$C for ~6 hours to improve metal-semiconductor contact quality and remove adsorbates from the surface of the substrate.

### Electrical and Optoelectronic Measurements

All electrical measurements were performed using a Keithley 4200-SCS parameter analyzer integrated with a probe station enclosed within a Faraday cage under ambient conditions. Spatially resolved photocurrent measurements were performed using a LABRAM Horiba system under continuous-wave 532 nm laser excitation and a x-y-z movement piezoelectric stage. The laser spot size was approximately 1 μm, and the spatial scanning step was approximately 0.1 μm. Transient photodetection measurements were performed using a Tektronix TBS1154 digital storage oscilloscope by monitoring the photocurrent response during illumination and after switching off the optical excitation. The rise time was defined as the time required for the photocurrent to increase from 10% to 90% following illumination, whereas the fall time was defined as the time required for the photocurrent to decrease from 90% to 10% after the illumination was switched off. The transient electrical signals were recorded using a digital oscilloscope interfaced with the Keithley measurement system. Wavelength-dependent photoresponse measurements were performed using THORLABs' 4-Wavelength High-Power LED Source with red (625 nm), green (505 nm), and blue (470 nm) illumination, using optical powers of 3.287, 9.334, and 10.45 mW, respectively. The corresponding illuminated spot diameter was approximately 2.5 mm.

Spatially resolved electroluminescence (EL) measurements were performed using an optical microscope equipped with a high-sensitivity, Peltier-cooled CCD camera from Andor Technology. A long-working-distance 100x objective was used to achieve high spatial resolution of the device, with an additional 2x magnification used with the microscope. The device was first located and focused in the electrically contacted configuration using tungsten probe tips and micromanipulators. A reference image of the device was acquired under microscope-lamp illumination with the camera gain set to zero. This reference image was subsequently overlaid with the EL image acquired at high camera gain to accurately identify the spatial origin of the emission within the active device region.

For spectrally resolved EL measurements, a Horiba LABRAM spectrometer equipped with a 600 grooves $mm^{-1}$ grating and a CCD detector was employed. The device was focused using a long-working-distance 50x objective while probing electrically. EL spectra were recorded under constant-current injection with an acquisition time of ~80 s for each spectral window. Multi-window spectral acquisition was performed over the photon-energy range of 1.1–2.8 eV. Time-resolved EL measurements were performed by recording the emission response under electrical excitation with temporal resolutions corresponding to frame intervals of 1 s and 0.01 s. Further temporal resolution was limited by the frame-acquisition rate capability of the CCD detector.

**Acknowledgement**

PS acknowledges the Department of Science and Technology (DST) (Project Codes: DST/NM/TUE/QM-1/2019; and National Quantum Mission (NQM) DST/QTC/NQM/QMD/2024/4/(G), India. TEM work was performed at the Sophisticated Analytical Technical Help Institute (SATHI), IIT Kharagpur, supported by DST, Govt of India. SC and BN acknowledge Mr. Arup Ghoshal for STEM measurement at SATHI. MS acknowledges the NNFC (National NanoFabrication Centre) at the CeNSE department, IISc, for the fabrication support.

## Author Contributions

PS and PR conceived the idea. PR performed the LHS synthesis, optical characterization, LHS device design, electro-optical measurements, and analyzed data. RV performed device fabrication, electro-optical measurements, and KPFM measurements, and designed the experiments. PR and RV performed the device characterization under the supervision of MS. SC, and BN contributed to CVD process development and performed STEM sample preparation and analysis. RV and BN supported the data analysis. PR and PS wrote the manuscript with input from RV, BN, MS, and SKV. PS supervised the project. All authors participated in discussing the results and reviewing the manuscript. # PR and RV contributed equally

## Additional information

Supplementary files available online

# Supplementary Information

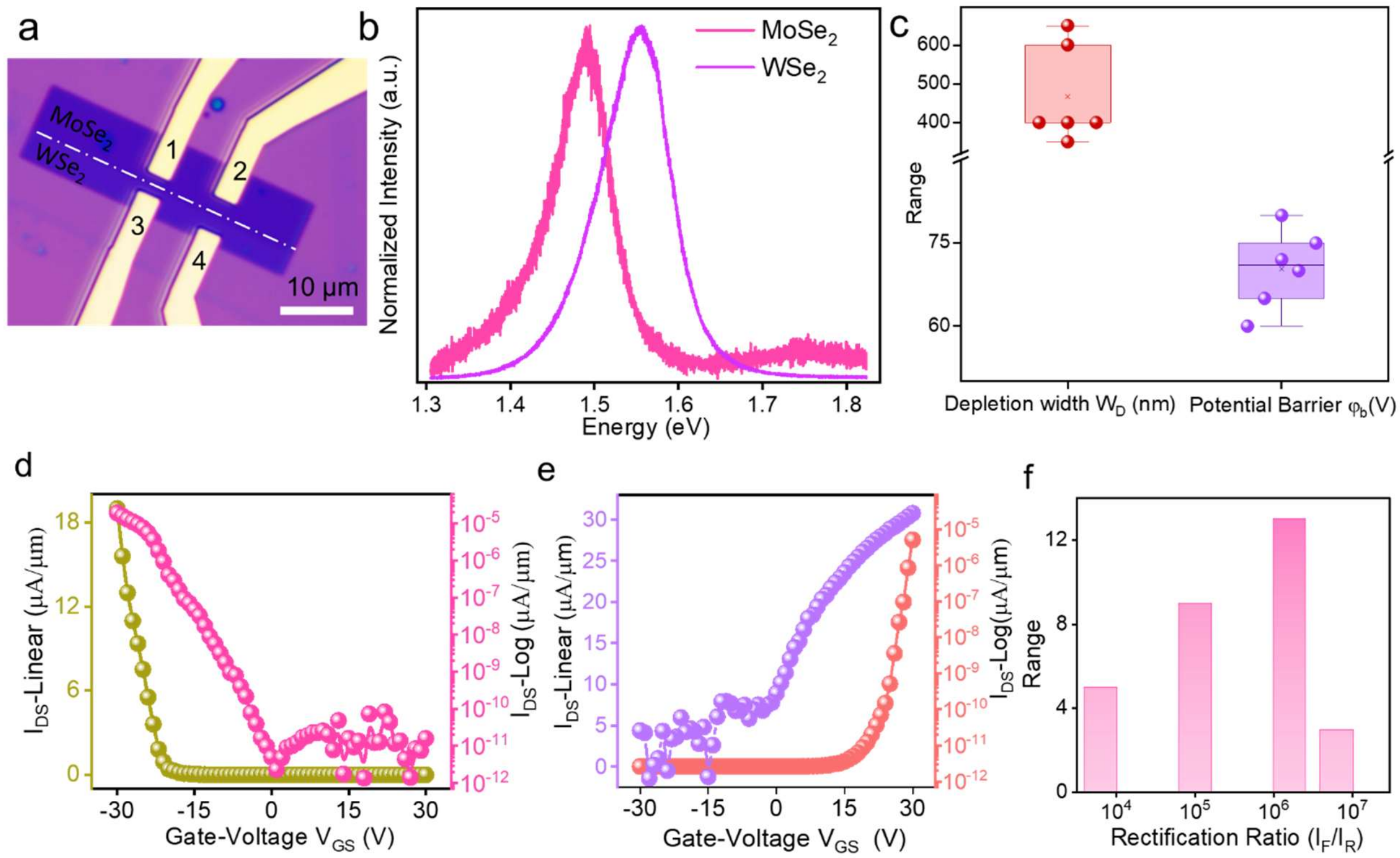

*Figure S1| Structural, optical and electrical characterization of $MoSe_2$-$WSe_2$ LHS. a: Optical microscopic image of the device showing the electrode configuration and measurement configuration used for electrical characterization. b: Normalized PL spectra acquired from $MoSe_2$ and $WSe_2$ domain, showing distinct excitonic emission peaks corresponding to each material c: Extracted depletion width and potential barrier from different set of devices. d,e: Transfer characteristics of $WSe_2$ and $MoSe_2$ domain respectively. f: Rectification ratio extracted from a set of devices, showing robust diode characteristics*

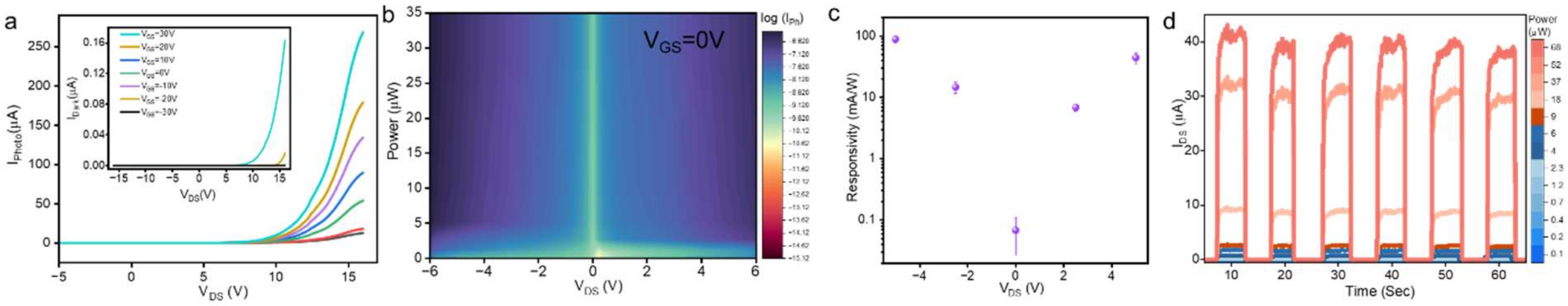

*Figure S2| Intrinsic photodetection characteristics of the LHS at $V_{GS}$=0V. a. $V_{GS}$-dependent $I_{DS}$–$V_{DS}$ characteristics measured under a fixed illumination power of 8 μW, demonstrating effective electrostatic control of the photo response. b. Photocurrent mapping shows at reverse bias the current is maximum c, Responsivity as a function of $V_{DS}$ under 30 mW power shows that at $V_{GS}$=0V $V_{DS}$=-4V shows higher responsivity compared to forward bias. d, Time-resolved $I_{PC}$ measured under periodic optical excitation, exhibiting stable and reproducible switching behavior.*

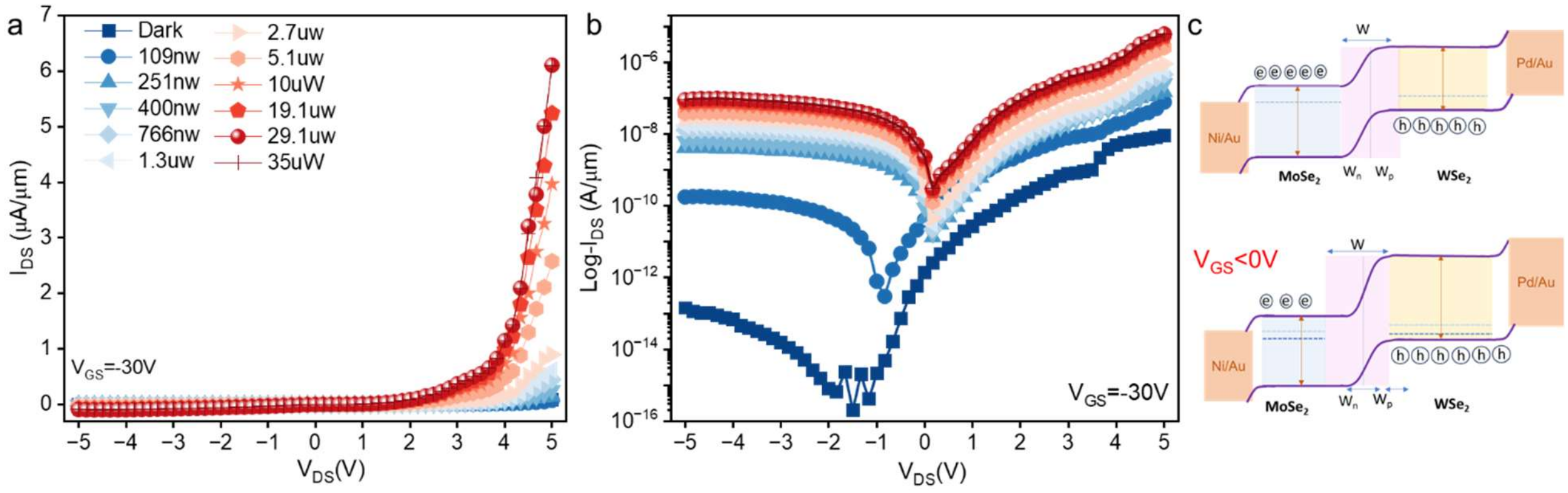

*Figure S3| Optical power dependent photodetection characteristics of the LHS at $V_{GS}$=-30V. a,b: **a,b:** Linear-scale (**a**) and logarithmic-scale (**b**) $I_{DS}$ vs $V_{DS}$ characteristics measured at $V_{GS}$=−30V under optical powers ranging from 0 to 35 μW. With increasing illumination power, the drain current increases systematically over the entire bias range due to the generation of a larger number of photocarriers. The logarithmic plot highlights the several-orders-of-magnitude enhancement in current under illumination while preserving the rectifying nature of the junction. **c:** Schematic band diagrams of the $WSe_2$–$MoSe_2$ heterojunction at $V_{GS}$=0 and $V_{GS}$<0V. Applying a negative $V_{GS}$ increases band bending across the junction, strengthening the built-in electric field and facilitating more efficient separation and collection of photogenerated carriers, which leads to the enhanced photocurrent observed in **a** and **b**.*

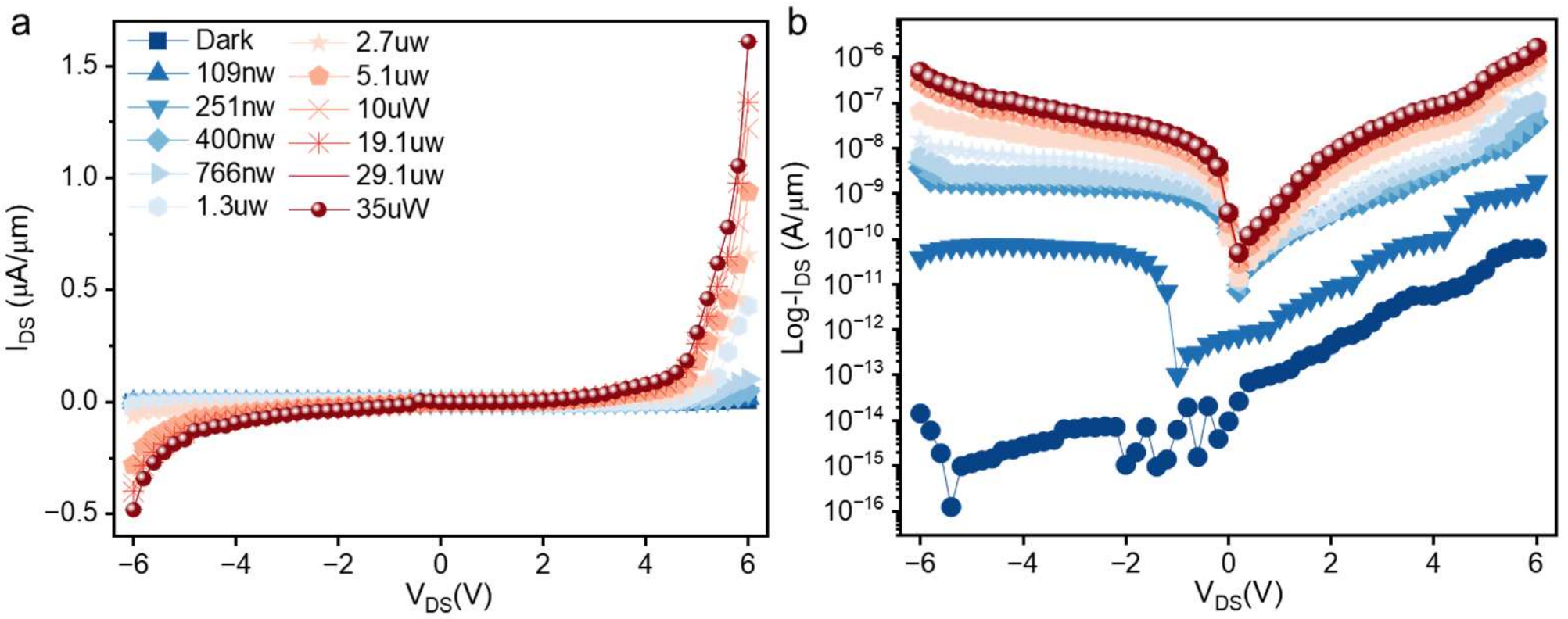

*Figure S4| Optical-power-dependent photoresponse at $V_{GS}$=+30V. a,b: Linear-scale (a) and logarithmic-scale (b) $I_{DS}$ $V_{DS}$ characteristics of the $WSe_2$–$MoSe_2$ heterojunction measured under optical powers ranging from dark to 35 μW at $V_{GS}$=+30V. The drain current increases progressively with increasing illumination power due to enhanced photocarrier generation. Compared to the $V_{GS}$=−30V condition, a much stronger increase in photocurrent is observed under reverse bias, as evident in both the linear and logarithmic plots.*

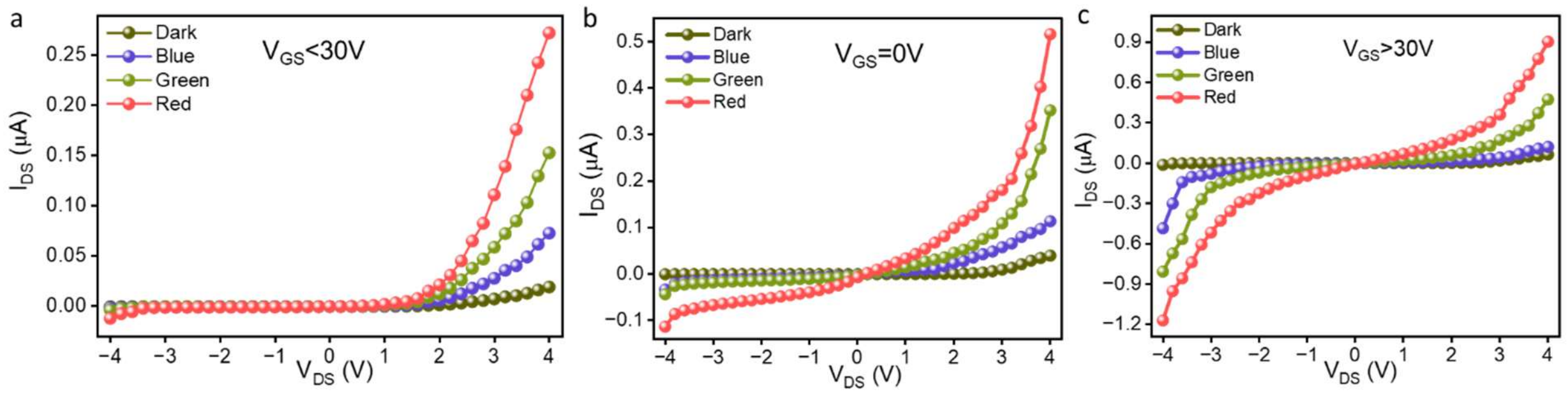

*Figure S5| Gate- and wavelength-dependent photodetection characteristics. a–c: $I_{DS}$–$V_{DS}$ characteristics measured under red, green, and blue LED illumination at different gate voltages. The photocurrent exhibits strong wavelength dependence, with longer-wavelength excitation producing the largest response. Gate modulation significantly alters the junction electrostatics, resulting in a transition from predominantly forward-bias photo-response at negative gate voltage to enhanced reverse-bias photodetection at positive gate voltage.*

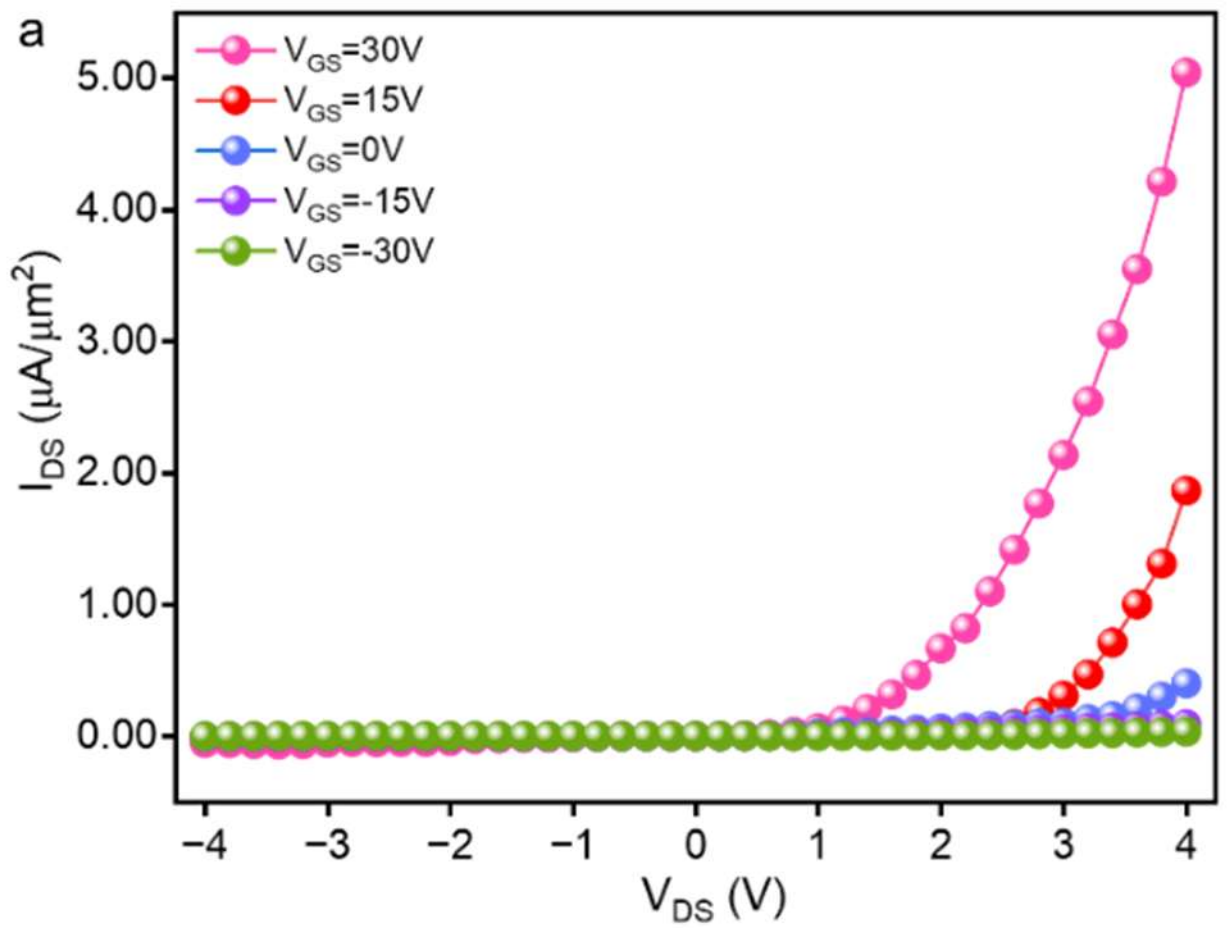

*Fig S6|Gate modulated photodetection characteristics under full device illumination condition at 10X magnification*

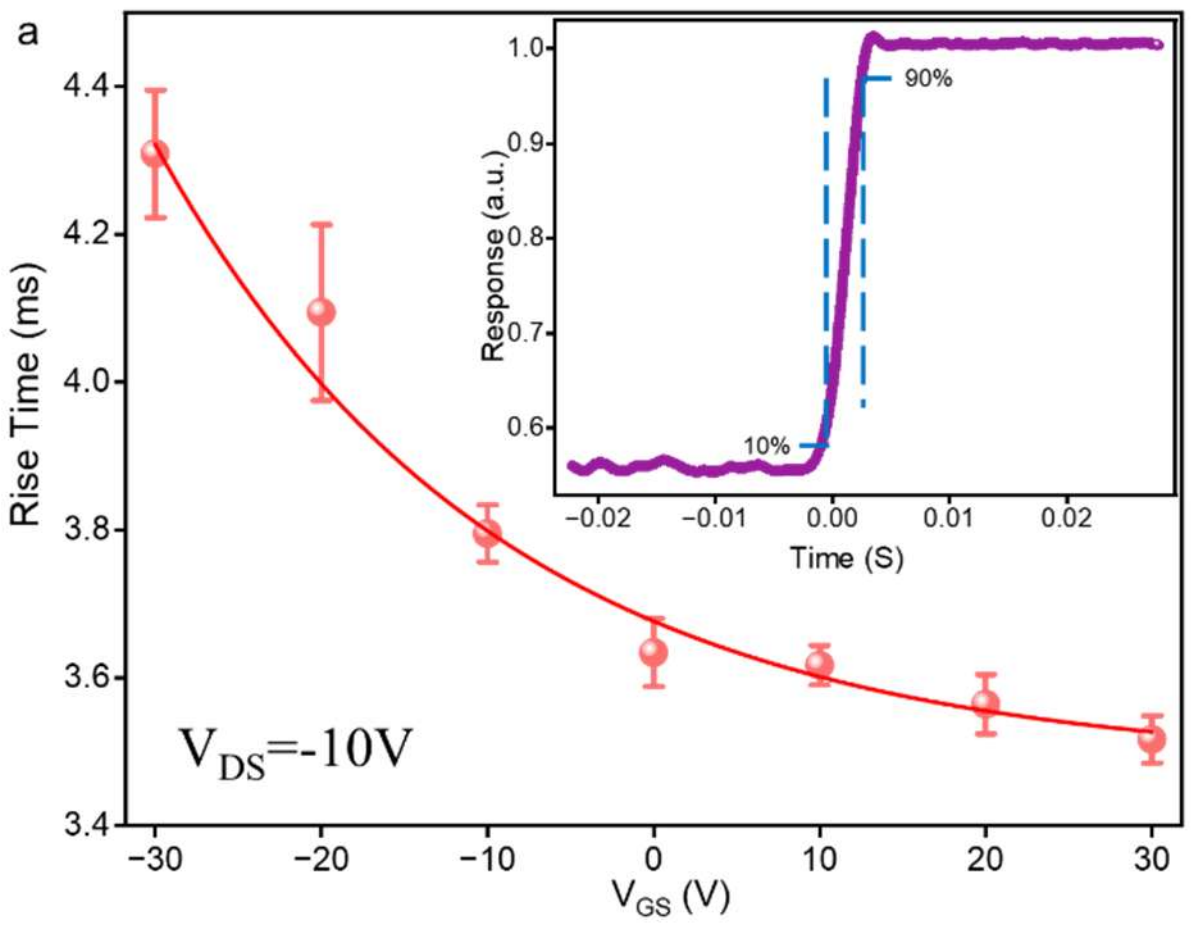

*Figure S7|Gate-dependent response speed (rise time) of the $WSe_2$–$MoSe_2$ heterojunction photodetector. a: Rise time ($\tau_r$) as a function of gate voltage ($V_{GS}$). The rise time decreases with increasing $V_{GS}$ , indicating faster photocarrier separation and collection under positive gate bias. Inset shows normalized transient photoresponse used to extract the rise time from the 10–90% transition region, marked by the dashed lines.*

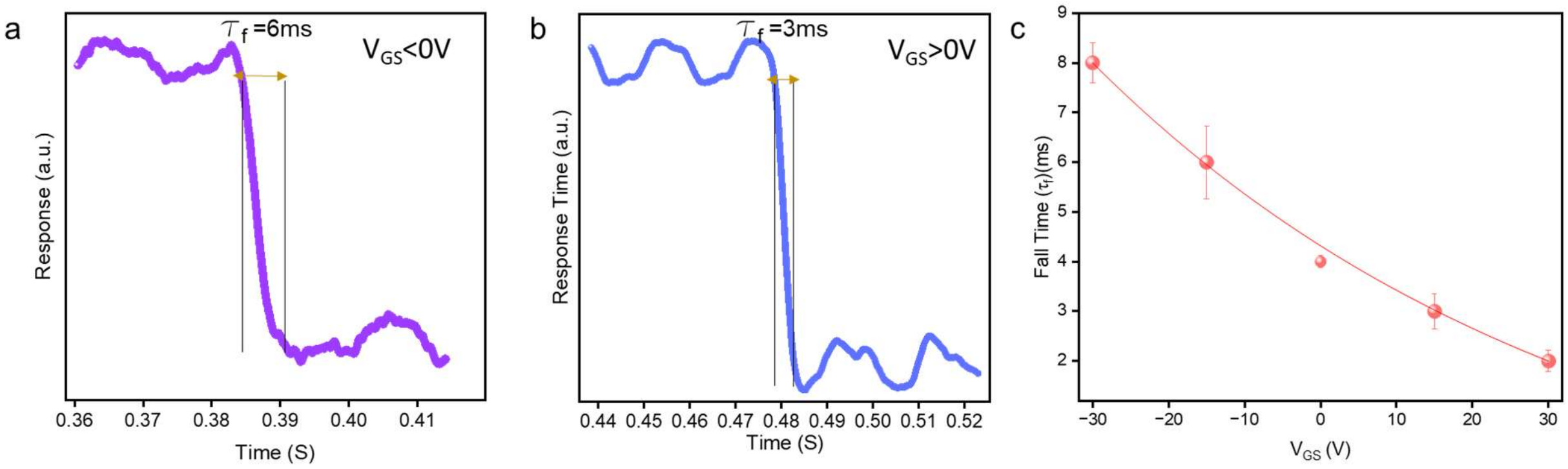

*Figure S8| Gate-dependent fall-time characteristics of the $WSe_2$–$MoSe_2$ heterojunction photodetector at $V_{DS}$=0. a,b, Representative transient photo response curves showing the decay of the photocurrent after the illumination is switched off at different gate voltages. The arrows indicate the extracted fall time ($\tau_f$) c, Extracted fall time as a function of gate voltage $V_{GS}$=0V. The fall time decreases progressively with increasing $V_{GS}$ , indicating faster carrier extraction and reduced carrier trapping under positive gate bias.*

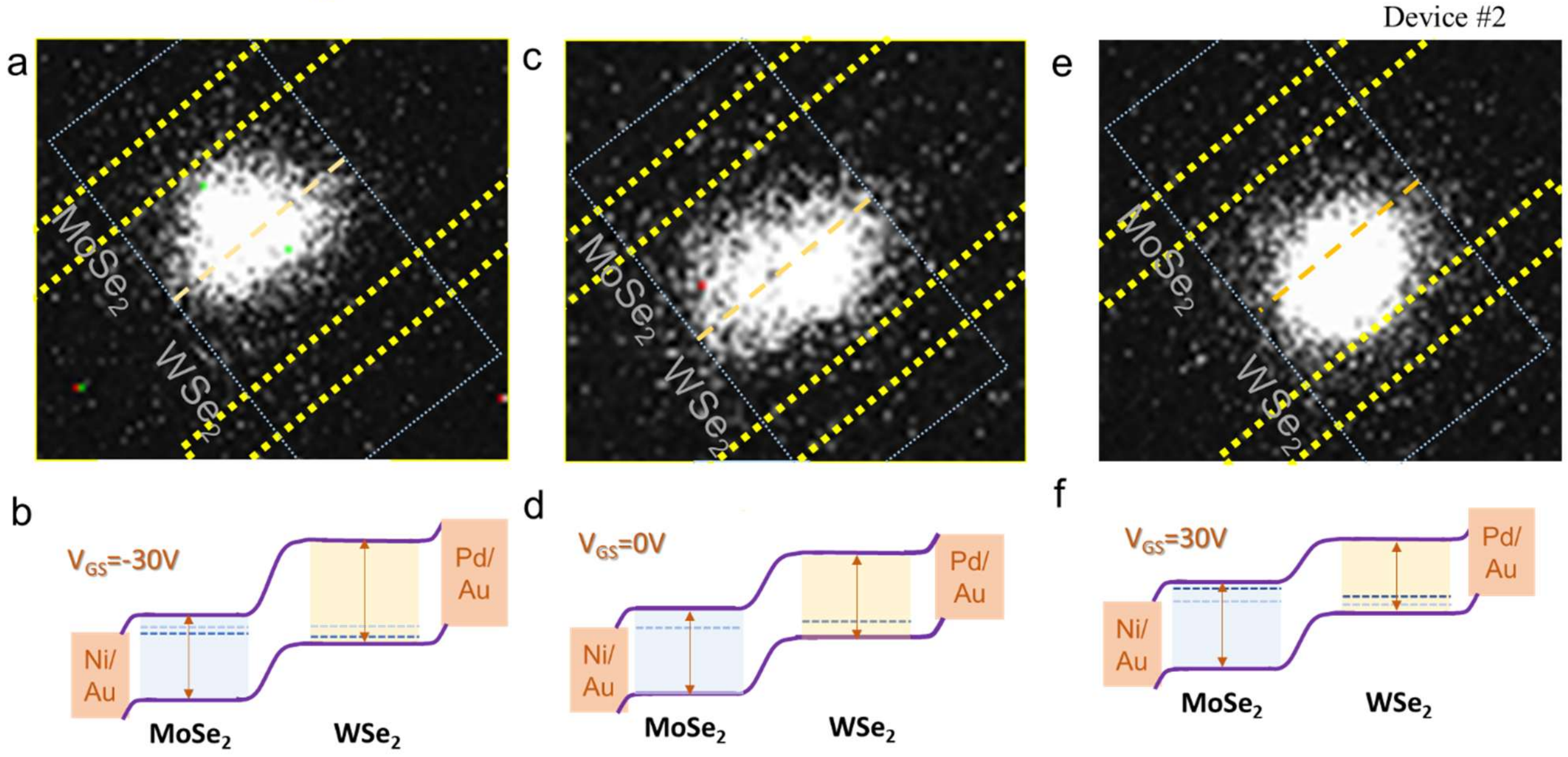

*Figure S9| Gate-controlled electroluminescence in the $WSe_2$–$MoSe_2$ lateral heterojunction. a,c & e, EL intensity maps measured at $V_{GS}$=-30: 0: +30V, respectively. The emission region shifts from the $MoSe_2$ side at negative gate bias to the heterointerface at $V_{GS}$=0V and further towards the $WSe_2$ side at positive gate bias. Bottom (b, d & f), Schematic band diagrams illustrating the gate-induced modulation of carrier injection and recombination, enabling electrostatic control over the EL emission location across the heterojunction.*

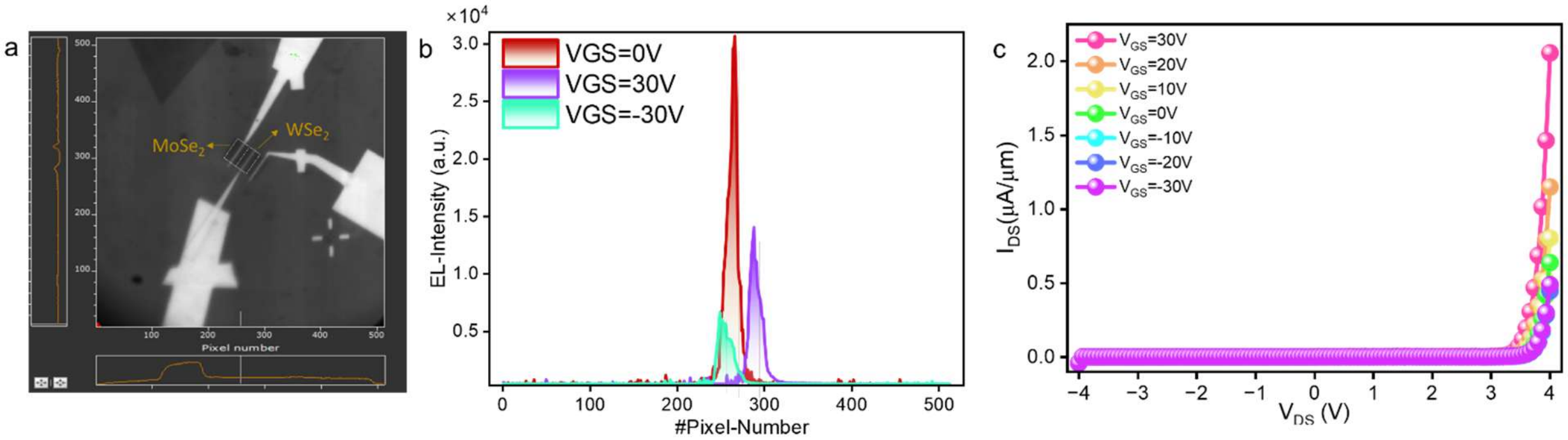

*Figure S10| Spatially resolved electroluminescence of the $WSe_2$–$MoSe_2$ heterojunction. a, Device image showing the line-scan direction across the heterointerface. b, EL intensity profiles measured under different gate voltages, revealing that the emission is strongest at the heterointerface and shifts across the junction with gate bias. c, Corresponding $I_{DS}$–$V_{DS}$ characteristics, showing that the enhanced EL intensity follows the increase in carrier injection current.*

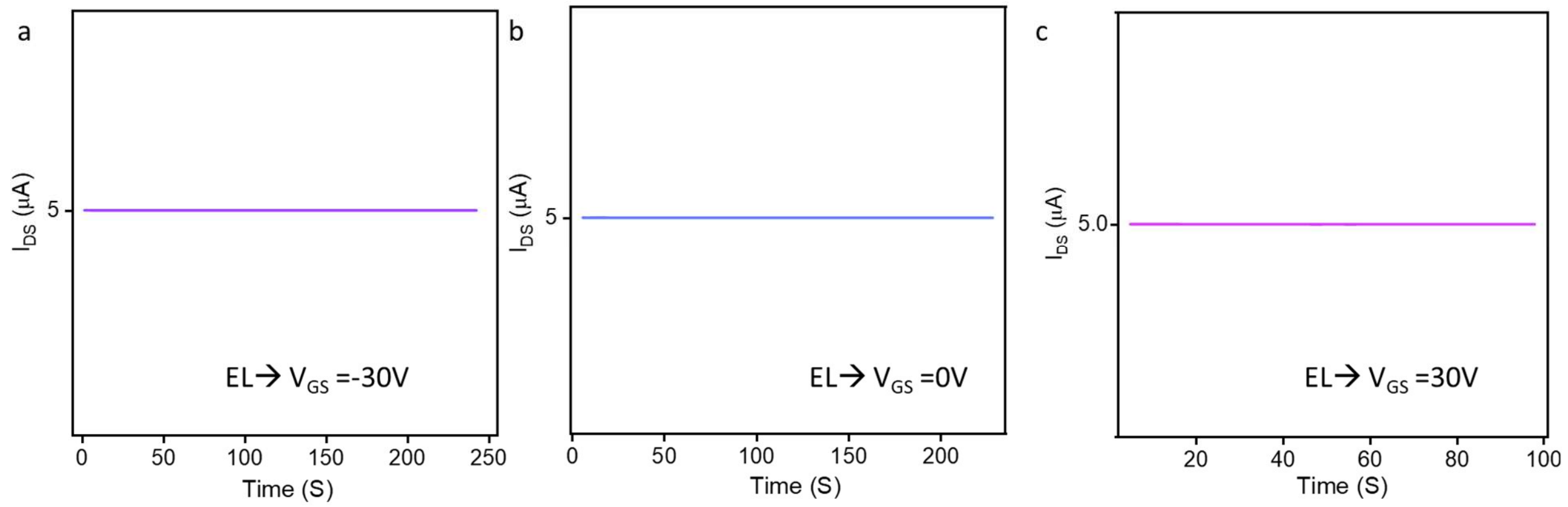

*Figure S11| Time-resolved electroluminescence at different gate voltages. a–c, EL intensity as a function of time under constant electrical injection. Stable emission is observed at negative and zero gate bias, whereas positive gate bias leads to a rapid increase and saturation of the EL intensity, indicating gate-enhanced carrier injection and radiative recombination.*

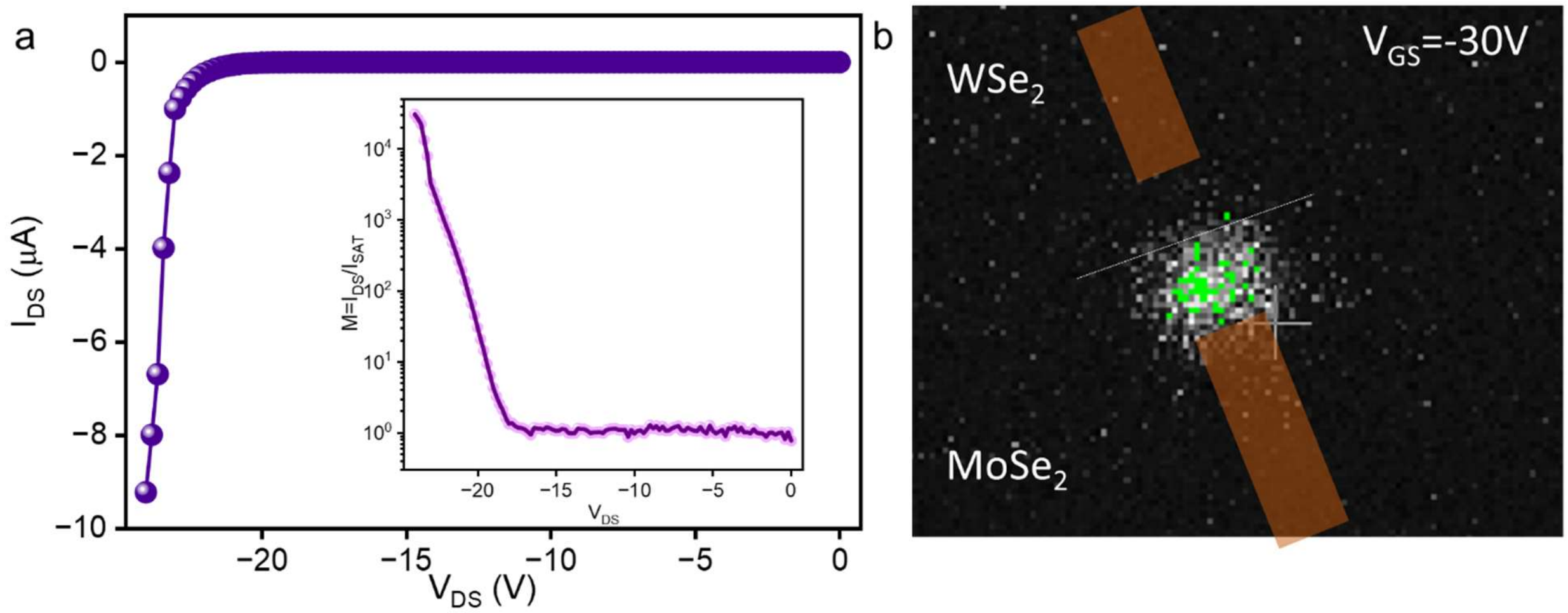

*Figure S12| Electroluminescence (EL) under reverse bias. a, Reverse-bias multiplication factor (M) and corresponding EL intensity, showing the onset of avalanche carrier multiplication. b, EL map at $V_{GS}$=−30V, indicating that the reverse-bias emission is localized at the heterojunction where impact ionization is strongest.*

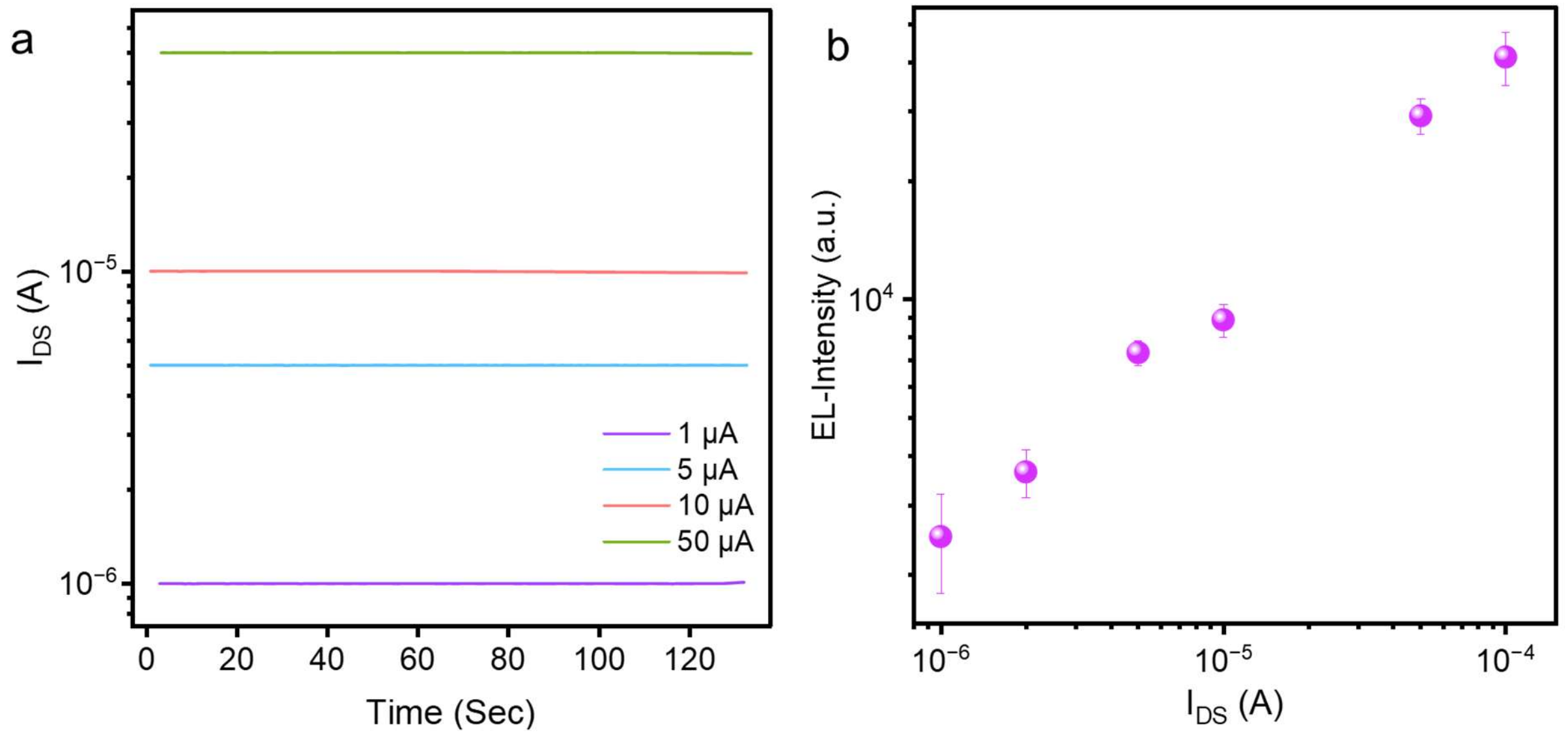

*Figure S13| Current-dependent electroluminescence. a, Time-dependent EL intensity under different constant current injection levels. b, Corresponding steady-state EL intensity, showing a monotonic increase with increasing injection current due to enhanced radiative recombination.*

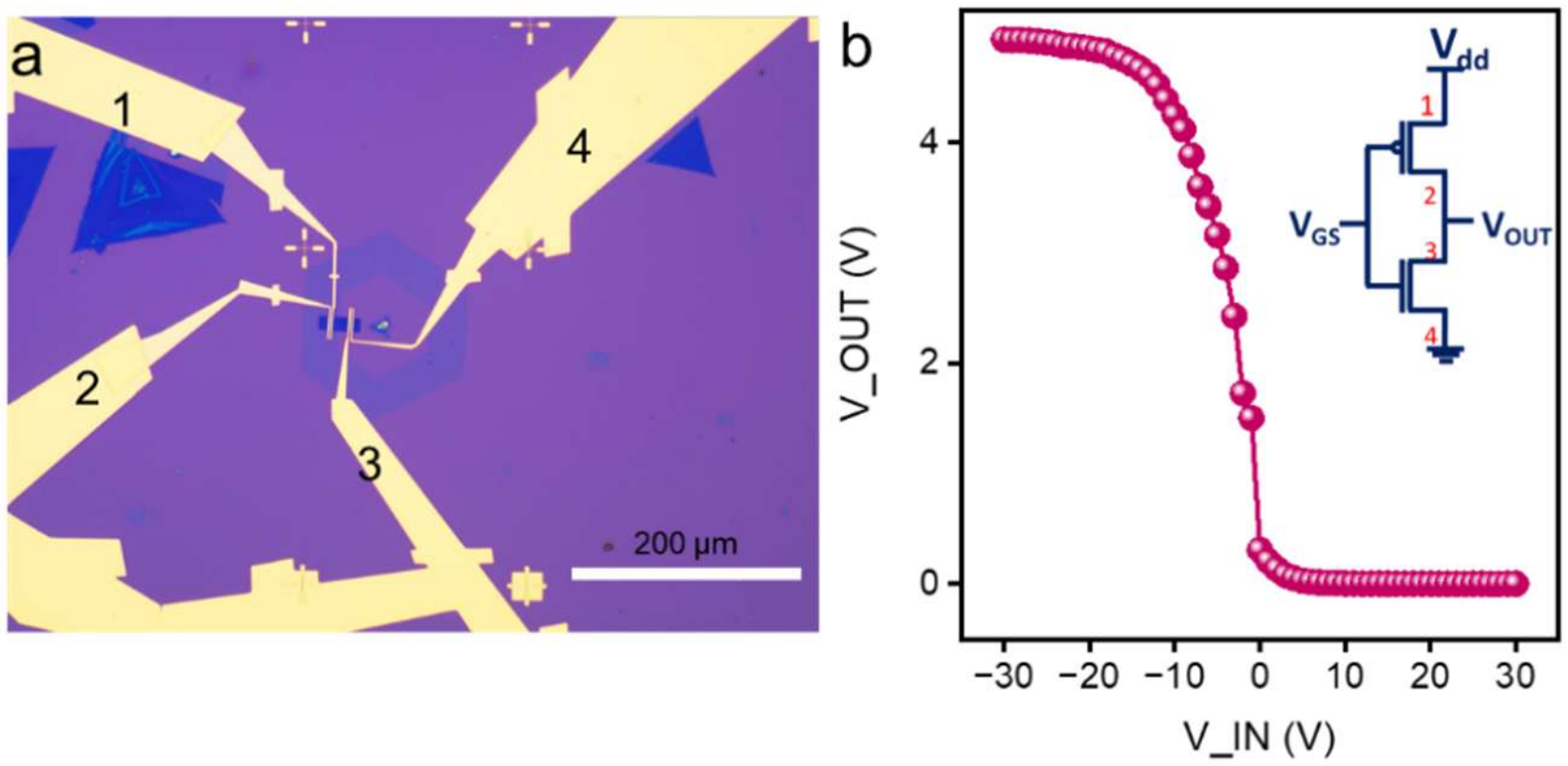

| Parameter | Value |
|---|---|
| $V_{OH}$ | ~5 V |
| $V_{OL}$ | ~0 V |
| $V_{IL}$ | ~−8 V |
| $V_{IH}$ | ~1 V |
| $NM_L$ | ~8 V |
| $NM_H$ | ~4 V |

*Figure S14|CMOS inverter based on a lateral $MoSe_2$–$WSe_2$ heterostructure.*
*a, Optical image of the inverter device and contact configuration. b, Voltage transfer characteristic showing a clear transition between logic states. Inset: circuit schematic of the CMOS inverter used for the measurements. c, extracted parameters from the VTC curve*

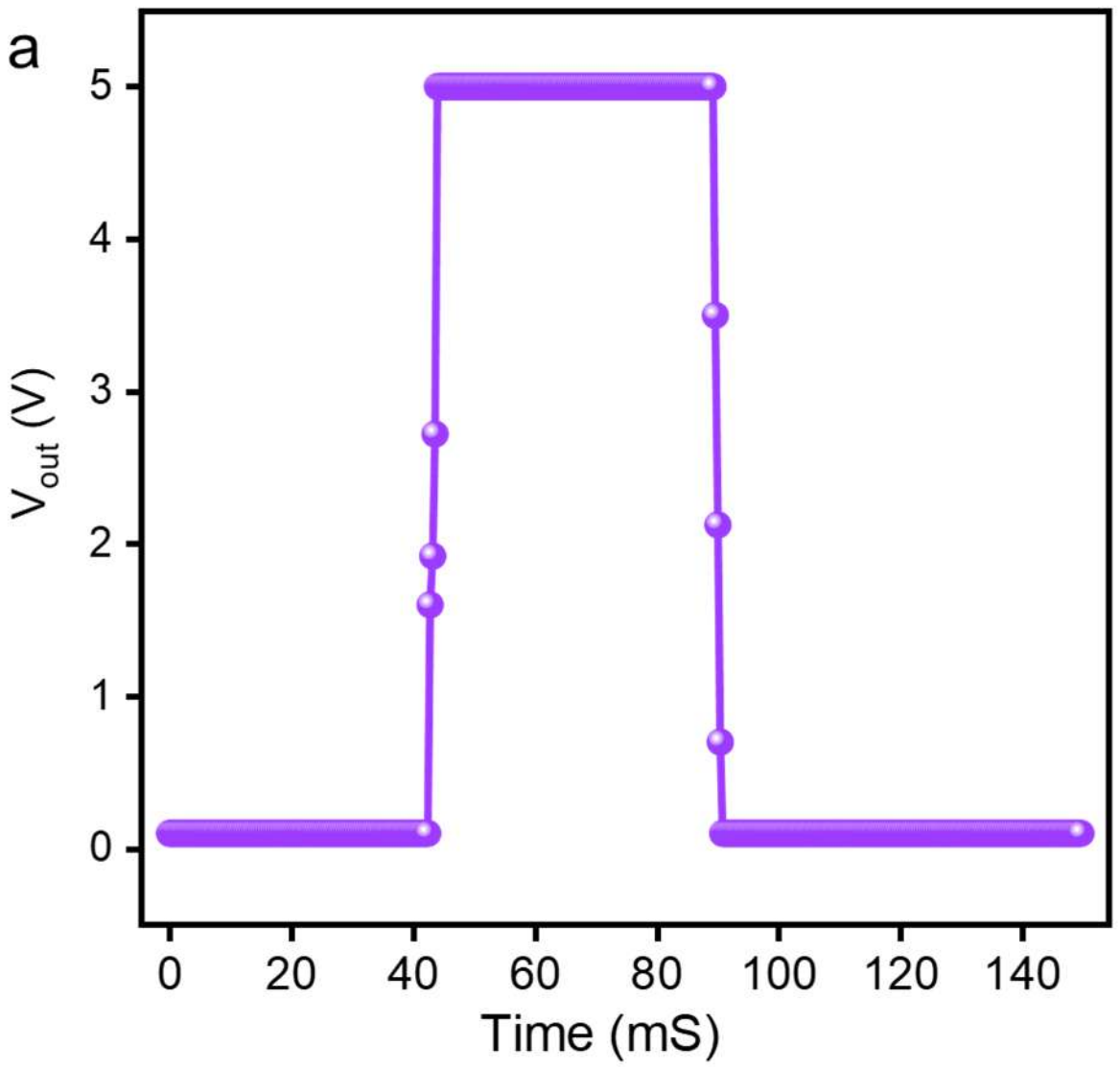

b

| Parameter | Value |
|---|---|
| HIGH State | ~5 V |
| LOW State | ~0.2 V |
| Voltage Swing | ~4.8 V |
| Rise Time | ~1–3 ms |
| Fall Time | ~1–2 ms |
| Propagation Delay | ~1–2 ms |
| Switching frequency | ~0.250 KHz |

*Figure S15| Dynamic performance of the CMOS inverter based on LHS. a: Output voltage waveform showing stable switching between HIGH and LOW logic states. b: extracted inverter parameters, including voltage swing, rise/fall times, propagation delay, operating frequency, and power consumption.*